\documentclass[a4paper,11pt]{article}
\pdfoutput=1 
\usepackage[utf8]{inputenc}
\usepackage{enumitem}  
\usepackage{jheppub} 
\usepackage[utf8]{inputenc} 
\usepackage[T1]{fontenc} 
\usepackage{array}
\usepackage{braket} 
\usepackage{siunitx}
\usepackage{subcaption}
\usepackage{multirow}
\usepackage[normalem]{ulem}

\newcommand{\cij}[2]{\ifnum#1=#2 C_{#1}^{c, 2} \else C_{#1}^c C_{#2}^c \fi}

\preprint{IPPP/19/48}
\title{\boldmath Charming New $B$-Physics}

\author[a]{Sebastian J{\"a}ger,
}
\author[b]{Matthew Kirk,}
\author[c]{Alexander Lenz}
\author[a]{and Kirsten Leslie}

\affiliation[a]{University of Sussex, Department of Physics and Astronomy,
  Falmer, Brighton BN1 9QH, UK}
\affiliation[b]{Dipartimento di Fisica, Università di Roma “La Sapienza” \& INFN Sezione di Roma, Piazzale Aldo Moro 2, 00185 Roma, Italy}
\affiliation[c]{IPPP, Department of Physics, Durham University, Durham DH1 3LE, UK}

\emailAdd{S.Jaeger@sussex.ac.uk}
\emailAdd{matthew.kirk@roma1.infn.it}
\emailAdd{alexander.lenz@durham.ac.uk}
\emailAdd{k.leslie@sussex.ac.uk}

\abstract{
We give a comprehensive account of the flavour physics of Beyond-Standard-Model (BSM) effects in $b \to c \bar{c} s$ transitions,
considering the full set of 20 four-quark operators. We discuss the leading-order structure of their RG mixing with
each other as well as the QCD-penguin, dipole, and FCNC semileptonic operators they necessarily mix with,
providing compact expressions. We also provide the first complete results for BSM effects in the lifetime
observables $\Delta \Gamma_s$ and $\tau(B_s)/\tau(B_d)$, as well as for the
semileptonic CP-asymmetry $a_{sl}^s$. From a global analysis, we obtain stringent constraints on 16 of the 20
BSM operators, including the  10 operators $Q^{c\prime }_{1 \dots 10}$ involving a right-handed strange quark.
Focussing on CP-conserving new physics, the constraints correspond to NP scales of order 10 TeV in most cases,
always dominated by exclusive and/or radiative $B$-decays via RGE mixing.
For the remaining four operators,  including the two Standard-Model (SM)
ones, larger effects are experimentally allowed, as previously noted in \cite{Jager:2017gal}.
We extend that paper's scope to
the CP-violating case, paying attention to the impact on the decay rate and time-dependent CP-violation in
$B_d \to J/\psi K_S$. Contrary to common lore, we show that quantifiable constraints arise for new physics
in either of the two SM operators, with the uncertain non-perturbative matrix element of
the colour-suppressed (or equivalently, colour-octet) operator determined from the data. For new physics in the coefficient $C^c_1$, suppressed in the SM, we find (in addition to CP-conserving new physics) two perfectly viable, narrow bands of
complex Wilson coefficients. Somewhat curiously, one of them contains a region where the fitted matrix element for
the colour-suppressed operator is in agreement with naive factorization, contrarily to a widely held belief that
large non-factorizable contributions to $B_d \to J/\psi K_S$ are implied by experimental data.

}

\begin{document} 
\maketitle
\flushbottom

\section{Introduction}
\label{sec:intro}
A wide range of  $B$ meson decays are affected by $b \to c \bar{c} s$ transitions, providing the opportunity
to use a rich set of complementary observables to detect possible new physics (NP), or place constraints on such
dynamics and its mass scale.
These partonic transitions are generated at the tree level in the SM, and contribute again at tree level to lifetime
observables such as \(\Delta \Gamma_s\) and \(\tau(B_s)/\tau(B_d)\) \cite{Bauer:2010dga}, which stand out among others through their
good theoretical control.
Moreover, as we have shown in \cite{Jager:2017gal}, there is the possibility that operators with this flavour structure
can be involved
in the rare $B$-decay anomalies \cite{Aaij:2014pli,Aaij:2015esa,Aaij:2019wad,Khachatryan:2015isa,Lees:2015ymt,Wei:2009zv,Aaltonen:2011ja,Aaij:2015oid,Abdesselam:2016llu,Wehle:2016yoi,Aaboud:2018krd,Sirunyan:2017dhj}\footnote{The possibility of virtual charm BSM physics in rare semileptonic decay was raised in \cite{Lyon:2014hpa}}. In this case, the contribution to the observables is through a charm
loop, which radiatively generates the semileptonic Wilson coefficient $C_{9V}$,
conjectured to lie behind the so-called $P_5'$ anomaly.\footnote{
As a peculiar feature, this mechanism can induce a $q^2$-dependent BSM contribution to
$C_{9V}$. Such a $q^2$ dependence, should it be implied by future experimental data,
might otherwise have been taken as an unambiguous sign for a hadronic origin of the anomaly.}
That BSM $b \to c \bar c s$ transitions can give important one-loop
contributions to rare radiative and semileptonic
decays should not really be
surprising, given the same happens in the SM, where charm loops provide on the order of one-half of the
$b \to s \gamma$ decay amplitude, as well as of $C_{9V}$. The size of these effects is amplified by strong
renormalization-group (RG) running in the SM, which can be even stronger for certain BSM operators \cite{Jager:2017gal}.
Neither does the NP scale need to be particularly low: because of strong RGE
running, accounting for the $P_5'$ anomaly merely requires a BSM contribution $|C_1^{\rm BSM}| \sim 0.1$, giving
a naive NP scale 
$$
   \Lambda_{\rm NP} \sim \left( \frac{4\, G_F}{\sqrt{2}} |V^*_{cs} V^{}_{cb}| \times 0.1 \right)^{-1/2} \sim \SI{3}{\TeV} \,.
$$ 
For a weakly coupled, sufficiently leptophobic tree-level mediator, this may be allowed by high-$p_T$ LHC searches, and
will not cause problems with electroweak precision observables. For strong coupling, the NP scale can be as high as \SI{30}{\TeV}
and out of the reach of LHC direct searches altogether \cite{Jager:2017gal}.
While the flavour anomalies provided some extra motivation for our earlier work \cite{Jager:2017gal}, it is clear that our BSM operators can only provide a lepton-flavour-universal contribution, and as such can only be a part of a more complete solution (see \cite{Geng:2017svp} and \cite{Alguero:2018nvb,Alguero:2019ptt,Aebischer:2019mlg} for discussion of flavour-universal and flavour-non-universal combinations).
Therefore, the scale suggested above is likely to be a lower bound if our operators contribute in this fashion.
In the rest of this paper however, we will be focussing our attention on the BSM effects arising from our full basis of operators in the observables already discussed.

A BSM model that generates new physics only in $b \to c \bar c s$ is not realistic, as is already evident from the leading-order
RG mixing we have mentioned.  Thus the model-independent study presented here
has to be considered as a building block in constructing or constraining UV models.

In the present paper, we extend the analysis of \cite{Jager:2017gal} in the following ways:
\begin{itemize}
\item We study a full basis of 20 $b \to c \bar c s$ operators, where \cite{Jager:2017gal} focused on those 4 that
can generate the $P_5'$ anomaly, and we obtain stringent constraints on the 16 others.
\item We extend the case of \(C^c_{1-4}\) to allow for CP-violating NP and show how to include $B_d \to J / \psi K_S$
data with minimal non-perturbative inputs.
\end{itemize}

To address the first item, we build the anomalous dimension matrix governing the RG evolution of the 34 relevant operators,
and review the all-order block structure.
We explain that, while some blocks first arise at one-loop order and others at two-loop order, an unambiguous leading-order
evolution operator arises, and obtain a compact numerical expression for it. This can be useful in understanding how
different four-quark operator coefficients affect the rare and semileptonic decays, but may also be useful to a reader
wishing to consider a particular UV model, particularly when multi-operator correlations are important that go beyond
the one- and two-parameter cases we consider in our model-independent phenomenology.

We also compute the complete BSM contribution to the $\Delta B=0$ and $\Delta B=2$ lifetime observables,
including the impact on the semileptonic CP asymmetry $a_{sl}^s$.

Regarding the second item, a key point is that once CP violation is switched on,  contributions to time-dependent
CP violation in $B_d \to J/\psi K_S$ are generated, the sine coefficient of which is precisely measured and usually taken
to provide a clean determination of the CKM angle $\beta$; this no longer holds in the present context. However, we find
that as long as NP only affects one of the two Wilson coefficients present in the SM, the global data set is sufficient
to determine the (complex) ratio of the two relevant non-perturbative matrix elements jointly with the complex Wilson
coefficient, such that theory input is needed only for the matrix element of the operator $Q^c_1$,
which on grounds of large-$N$ arguments alone should be close to its naive-factorization value
(as is also borne out by
QCD factorization \cite{Beneke:1999br,Beneke:2000ry,Beneke:2001ev}, even though QCDF is not expected to provide reliable quantitative results -- see our main discussion in Section~\ref{sub:BtoJpsiK}).

As phenomenological results, for the 16 operators not previously considered we provide constraints for the CP-conserving
case, almost all of which turn out to be of order 10 TeV. In addition, we discuss in detail the constraints on the two one-parameter
CP-violating scenarios where we have a complex BSM contribution to one of the two SM coefficients $C^c_{1,2}$.
As a perhaps curious result, we find that among the allowed regions is one where CP-violating
BSM affects $C_1$ in such a way that the matrix element of $Q^c_2$ is \textit{also} close to its naive-factorization value.

We note that BSM effects in hadronic tree-level decays have previously been systematically studied in \cite{Bobeth:2014rda,Brod:2014bfa}. In these works all hadronic decay channels of the $b$ quark were considered and the new contributions to the Wilson coefficients were also allowed to be complex, but only the SM operators $Q^c_1$ and $Q^c_2$ were investigated.

The layout of our paper is as follows:
In Section \ref{sec:setup} we describe the setup of our model, specifying our operator basis and giving the
RG evolution. Section \ref{sec:observables} is devoted to our theoretical results for the lifetime observables, radiative and rare decay contributions, as well as our approach to $B_d \to J/\psi K_S$, and constitutes one of two main parts of our paper.
As the second main part, we describe our phenomenological constraints in Section \ref{sec:pheno} (details of the inputs we use are is given in Section \ref{sub:inputs}). Our conclusions can be found in Section \ref{sec:conclusions}.

\section{Setup}
\label{sec:setup}

\subsection{Operator basis}

For our study of BSM $b \to c \bar{c} s$ effects we use the following weak effective Hamiltonian
\begin{equation}
  \mathcal{H}_{\mathrm{eff}}=
  \mathcal{H}^{c\bar{c}}_{\mathrm{eff}}
  +\mathcal{H}^{rsl}_{\mathrm{eff}}
  +\mathcal{H}^{QCD}_{\mathrm{eff}}\, .
\label{Heff}
\end{equation}
The first part $\mathcal{H}^{c\bar{c}}_{\mathrm{eff}}$ contains a complete set of
 20 $b \to c \bar c s$ four-fermion operators, reading
\begin{equation}
  \mathcal{H}^{c\bar{c}}_{\mathrm{eff}}=\frac{4G_F}{\sqrt{2}}
   \lambda_c\sum\limits_{i=1}^{10} \left[ C^c_i(\mu)Q^c_i(\mu)+C_i^{c\prime }(\mu)Q_i^{c\prime }(\mu)\right] +   \mathrm{h.c.} \, \, .
   \label{eq:Heff_bscc}
\end{equation}
Here $\lambda_p=V_{pb}^{}V^*_{ps}$ for $p=u,c,t$ are CKM structures,
and the Wilson coefficients  $C^{c (\prime)}_i$ and operators $Q_i^{c(\prime)}$ are 
renormalized at the scale $\mu$. The operators $Q_i^c$ read
\begin{eqnarray}\label{eq:ccbar}
Q_1^c &=& (\bar c_L^i \gamma_\mu b_L^j) (\bar s_L^j \gamma^\mu c_L^i) , 
\qquad \quad
Q_2^c = (\bar c_L^i \gamma_\mu b_L^i) (\bar s_L^j \gamma^\mu c_L^j) ,
\nonumber \\[2mm]
Q_3^c &=& (\bar c_R^i b_L^j) (\bar s_L^j c_R^i) ,
\qquad \qquad \quad
Q_4^c = (\bar c_R^i  b_L^i) (\bar s_L^j  c_R^j) ,
\nonumber \\[2mm]
Q_5^c &=& (\bar c_R^i \gamma_\mu b_R^j) (\bar s_L^j \gamma^\mu c_L^i) , 
\qquad \quad
Q_6^c = (\bar c_R^i \gamma_\mu b_R^i) (\bar s_L^j \gamma^\mu c_L^j) ,
\nonumber \\[2mm]
Q_7^c &=& (\bar c_L^i b_R^j) (\bar s_L^j c_R^i) ,
\qquad \qquad \quad
Q_8^c = (\bar c_L^i  b_R^i) (\bar s_L^j  c_R^j) ,
\nonumber \\[2mm]
Q_9^c &=& (\bar c_L^i \sigma_{\mu\nu} b_R^j) (\bar s_L^j \sigma^{\mu\nu} c_R^i) ,
\qquad 
Q_{10}^c = (\bar c_L^i  \sigma_{\mu\nu}  b_R^i) (\bar s_L^j \sigma^{\mu\nu} c_R^j) ,
\label{eq:bscc_operator_basis}
\end{eqnarray}
where $i,j$ are $SU(3)$ colour indices 
and $\psi_{L,R}$ denotes the projections  $\psi_{L,R}=(1\mp\gamma^5)/2 \cdot \psi$.
The remaining 10 operators $Q_i^{c\prime}$ are obtained from those displayed
by letting ${L/R}\rightarrow {R/L}$.
In the SM the operators $Q_1^c$ and $Q_2^c$ arise due to a tree-level $W$ exchange. All other
18 operators are genuine BSM effects. For the overall normalisation of these operators we have chosen the 
CKM normalisation of the SM operators, which contains a tiny imaginary part.

For radiative and semileptonic decays $b \to s \gamma$ and $b \to s \ell^+ \ell^-$, further operators contribute.  These 
comprise the radiative and semileptonic operators
\begin{equation}
Q_{7\gamma} =  \frac{em_b}{16\pi^{2}}(\bar s_L \sigma_{\mu\nu}b_R) F^{\mu\nu},
\qquad
 Q_{9V} =  \frac{\alpha}{4\pi}(\bar s_L \gamma_\mu b_L) (\bar \ell\gamma^\mu \ell)\, ,
\end{equation}
and their parity conjugates, again denoted by primes. $m_{b}$ denotes the $\overline{\rm MS}$ b-quark mass,
$e$ is the electromagnetic coupling,
and $\alpha=\frac{e^{2}}{4\pi}$ . These operators enter through the
second part $\mathcal{H}^{rsl}_{\mathrm{eff}}$ of \eqref{Heff}
\begin{equation}
  \mathcal{H}^{rsl}_{\mathrm{eff}}=
  - \frac{4G_F}{\sqrt{2}} \lambda_t
  \left[ C_{7\gamma}(\mu)Q_{7\gamma}(\mu)+C_{9V}(\mu)Q_{9V}(\mu)+C^{\prime}_{7\gamma}(\mu)Q^{\prime}_{7\gamma}(\mu)+C^{\prime}_{9V}(\mu)Q^{\prime}_{9V}(\mu)
    \right]+ \mathrm{h.c.}  \, \, .
 \end{equation}
We note for the reader that since we do not have any effects in the \(Q^{(')}_{10}\) operators, the decay \(B_s \to \mu \mu\) is purely SM-like in our scenario.

Moreover, they receive contributions from QCD penguin operators, contained in the third term $\mathcal{H}^{QCD}_{\mathrm{eff}}$ of \eqref{Heff},
\begin{equation}
  \mathcal{H}^{QCD}_{\mathrm{eff}}= - \frac{4G_F}{\sqrt{2}} \lambda_t \left\{
    \sum\limits_{i=3}^{6} \left[C^p_{i}(\mu)P_{i}(\mu) + C^{p\prime}_i(\mu) P_i^{\prime}(\mu) \right]
    +C_{8g}(\mu)Q_{8g}(\mu)   + C_{8g}^{\prime}(\mu) Q_{8g}^{\prime}(\mu) \right\} \, .
\end{equation} 
Our QCD penguin operators are defined in \cite{Chetyrkin:1996vx} and include the chromomagnetic dipole operator,
reading explicitly:
\begin{eqnarray}
  {P}_3&=&(\bar{s}_L\gamma_{\mu}b_L)\sum\limits_q(\bar{q}\gamma^{\mu}q) \, ,
  \qquad \qquad \quad
  {P}_4=(\bar{s}_L\gamma_{\mu}T^ab_L)\sum\limits_q(\bar{q}\gamma^{\mu}T^aq) \, ,
  \nonumber \\
  {P}_5&=&(\bar{s}_L\gamma_{\nu}\gamma_{\mu}\gamma_{\rho}b_L)\sum\limits_q(\bar{q}\gamma^{\nu}\gamma^{\mu}\gamma^{\rho}q) \, ,
   \quad
   {P}_6=(\bar{s}_L\gamma_{\nu}\gamma_{\mu}\gamma_{\rho}T^ab_L)\sum\limits_q(\bar{q}\gamma^{\nu}\gamma^{\mu}\gamma^{\rho}T^aq) \, ,
   \nonumber \\
Q_{8g}   & =& \frac{g_sm_b}{16\pi^2}(\bar{s}\sigma^{\mu\nu}T^{a}P_Rb)G^a_{\mu\nu} \, ,
\end{eqnarray}
where $q$ runs over all active quark flavours in the effective theory, i.e. $q = u,d,s,c,b$, and a prime
again denotes a chirality conjugate. 
Our complete operator basis comprises 34 operators and closes under QCD renormalization.
We renormalize our operators as in \cite{Grinstein:1987vj,Grinstein:1990tj,Misiak:1991dj,Chetyrkin:1996vx}.
We will neglect the tiny contribution $|\lambda_u| \approx 0.00084$\footnote{Considering CP violating
  observables this approximation might be violated, since $\Im (\lambda_u)$ is of a similar size
  as  $\Im (\lambda_{c,t})$, see the discussion at the end of Section \ref{sub:Bmix}.} 
\cite{Charles:2004jd}
(for similar results see \cite{Bona:2006ah})
and we thus get from unitarity of the CKM matrix $\lambda_c = - \lambda_t$.

To isolate the BSM contribution, we split the Wilson coefficients
into SM parts and BSM parts,
\begin{align}
C_i^c(\mu)&=
      C_i^{\rm SM}(\mu)+\Delta C_i(\mu) \, .
\end{align}%
We will neglect the small mass ratios $m_{q}/m_W$ and $m_s/m_b$, which implies the vanishing of all primed
Wilson coefficients in the SM, $C_i^{\prime, {\rm SM}}(\mu) = 0$ and in addition the vanishing of the four-quark
coefficients $C_i^{c, {\rm SM}}(\mu) = 0$ for $i \not= 2$.
For our phenomenology we will also assume that at an input scale $\mu_0 \sim M_W$, the $\Delta C_i$ vanishes
for all but the four-quark operators. This corresponds to what we called `Charming Beyond the Standard Model' (CBSM) scenario in
\cite{Jager:2017gal}. As described below, RG evolution then generates BSM contributions to
the penguin and dipole operators, which play a crucial role in the phenomenology of the CBSM scenario. The CBSM scenario should
be viewed as a partial effective description of a more complete UV scenario, which will in general also involve nonzero initial values
for the other $\Delta C_i$.

\subsection{Renormalization-group evolution}
\label{sub:rge}
As emphasized in our previous work \cite{Jager:2017gal}, operator mixing can have a dramatic impact on the radiative and
semileptonic Wilson coefficients, and as a result on the contributions of the $b \to c \bar c s$ operators to the
radiative and rare semileptonic decays. It is therefore crucial to include its effects, which is conveniently done
through renormalization-group evolution from the BSM scale $\mu_0$ where the Wilson coefficients are initially
obtained to a scale $\mu \sim m_b$ appropriate to evaluating $B$-physics observables.

Collecting the 17 unprimed Wilson coefficients into a vector $\vec C$, the coupled system of renormalization-group equations
(RGE) governing their dependence on the renormalization scale $\mu$ can be written in matrix notation as
\begin{equation}
\mu\frac{d}{d\mu}\vec{C}(\mu)=\gamma^T(\mu) \vec{C}(\mu)\, ,  \label{eq:rge}
\end{equation}
where $\gamma$ is the anomalous-dimension matrix. Note that the unprimed and primed Wilson coefficients do
not mix at any order in QCD; moreover, the 17 primed Wilson coefficients, collectively denoted $\vec C^{\prime}$, fulfil
the same set of equations \eqref{eq:rge}, with identical anomalous-dimension matrix. Both statements follow directly from the parity invariance of QCD.
The solution of \eqref{eq:rge} is a linear relation
\begin{equation}
\vec C(\mu) = U(\mu, \mu_0) \vec C(\mu_0)  \label{eq:rgesol}
\end{equation}
in terms of the so-called evolution operator $U(\mu, \mu_0)$, which itself satisfies \eqref{eq:rge}.
Importantly, the SM and BSM parts $\vec C^{\rm SM}$ and $\Delta \vec C$ separately satisfy \eqref{eq:rgesol}.

If the leading-order anomalous-dimension matrix has the form
\begin{equation}
\gamma(\mu)=\frac{\alpha_s(\mu)}{4\pi}\gamma^{(0)} \, ,  \label{eq:admlo}
\end{equation}
where $\gamma^{(0)}$ is a constant matrix, then the explicit form of the evolution operator to leading logarithmic
order is
\begin{equation}  \label{eq:rgesollo}
  U^{(0)}(\mu,\mu_0)= \exp\left[-\frac{(\gamma^{(0)})^T}{2\beta_0}\ln\left(\frac{\alpha_s(\mu)}{\alpha_s(\mu_0)}\right)\right]
  \, ,
\end{equation}
where $\beta_0$ is the leading coefficient in the QCD $\beta$-function,

\begin{equation}
  \beta(\alpha_s) = \mu \frac{d\alpha_s}{d\mu} = - 2 \beta_0 \frac{\alpha_s^2}{4 \pi} + \dots
 \, .
\end{equation}
In the CBSM setup, two sources of complication arise. The first is that the leading-order
anomalous-dimension matrix does not have the form \eqref{eq:admlo}. The second is that part of the
anomalous-dimension matrix first arises at 2-loop order.
To discuss both issues further, let us decompose the Wilson coefficient vector $\vec C$ into subcomponents,
\begin{equation}\label{eq:cvec}
\vec{C}(\mu)=(\vec{C}^c(\mu),\vec{C}^p(\mu),C_{8g}(\mu),C_{7\gamma}(\mu),C_{9V}(\mu)) .\\
\end{equation}
The anomalous-dimension matrix then has the block form
\begin{equation}   \label{eq:admblock}
\gamma=\left(
\begin{array}{ccccc} 
\hat{\gamma}_{cc}&\hat{\gamma}_{cp}&\vec{\gamma}_{c8}&\vec{\gamma}_{c7}&\vec{\gamma}_{c9}\\
0&\hat{\gamma}_{pp}&\vec{\gamma}_{p8}&\vec{\gamma}_{p7}&\vec{\gamma}_{p9}\\
0&0&\gamma_{88}&\gamma_{87}&0\\
0&0&0 &\gamma_{77}&0\\
0&0&0&0 & \gamma_{99} \\ 
 \end{array}
\right) \, ,
\end{equation}
which holds to all orders of QCD.\footnote{The $(7c)$, $(8c)$, $(7p)$, $(8p)$,
$(79)$, and $(89)$ entries vanish in a massless scheme
due to the different dimensionality of the dipole and four-fermion operators,
 the $(pc)$ block because of the flavour symmetries of massless QCD.
} The element $\gamma_{99}$ vanishes for our normalization of $Q_{9V}$, which makes it proportional to a conserved quark current, but
we will consider a different normalization shortly.
$\gamma_{77}$, $\gamma_{88}$,  $\vec \gamma_{cp}$, and $\hat \gamma_{cc}$ first arise
at one-loop order, giving rise to the form \eqref{eq:admlo} at leading order. $\vec \gamma_{p7}$ and
$\vec \gamma_{p8}$ arises first at two-loop order. This makes them scheme-dependent already at leading order, though
the explicit coupling factors in the definitions of $Q_{7\gamma}$ and $Q_{8g}$ imply that
the form \eqref{eq:admlo} is maintained. On the other hand,
some of the elements of $\vec \gamma_{c7}$, $\vec \gamma_{c8}$, $\vec \gamma_{c9}$ and $\vec \gamma_{p9}$
are non-zero at one loop, but with no accompanying factor of $\alpha_s$.

To make this more explicit, let us
decompose the set of $(\bar c b) (\bar s c)$ Wilson coefficients further, $\vec C = (\vec C_{1 \dots 6}, \vec C_{7 \dots 10})
\equiv (\vec C_A, \vec C_B)$,
and further decompose the anomalous dimension accordingly,
\begin{equation} \label{eq:adm}
\gamma=\left(
\begin{array}{cccccc} 
\hat{\gamma}_{AA} & 0 & \hat{\gamma}_{Ap} & \vec{\gamma}_{A8} & \vec{\gamma}_{A7} & \vec{\gamma}_{A9} \\
0 & \hat{\gamma}_{BB} & 0 & \vec{\gamma}_{B8} & \vec{\gamma}_{B7} & 0 \\
0 & 0 & \hat{\gamma}_{pp} & \vec{\gamma}_{p8} & \vec{\gamma}_{p7} & \vec{\gamma}_{p9} \\
0 & 0 & 0 & \gamma_{88} & \gamma_{87} & 0 \\
0 & 0 & 0 & 0 & \gamma_{77} & 0 \\
0 & 0 & 0 & 0 & 0 & \gamma_{99} \\ 
 \end{array}
\right) \,  
\stackrel{\rm LO}{\sim} \left(
\begin{array}{cccccc} 
\frac{\alpha_s}{4\pi} & 0 & \frac{\alpha_s}{4\pi} &\frac{\alpha_s}{4\pi} & \frac{\alpha_s}{4\pi} &1 \\
0 & \frac{\alpha_s}{4\pi}  & 0  & 1 & 1 & 0 \\
0 & 0 & \frac{\alpha_s}{4\pi} & \frac{\alpha_s}{4\pi} & \frac{\alpha_s}{4\pi} & \frac{\alpha_s}{4\pi} \\
0 & 0 & 0 & \frac{\alpha_s}{4\pi} & \frac{\alpha_s}{4\pi} & 0 \\
0 & 0 & 0 & 0 & \frac{\alpha_s}{4\pi} & 0 \\
0 & 0 & 0 & 0 & 0 & \frac{\alpha_s}{4\pi}  \\ 
 \end{array}
\right) .
\end{equation}
In the second expression we have indicated in what blocks factors of $\alpha_s/(4 \pi)$ do or do not arise at leading order.
The loop- and coupling-order of the leading contributions to each block are also given in Table \ref{tab:looporders}
(the full algebraic expressions for each block are given in Appendix~\ref{app:adm}).
In (\ref{eq:adm}) we have also made explicit which blocks vanish to all orders in QCD. Besides those blocks whose vanishing is
already expressed through (\ref{eq:admblock}), these are the blocks concerning the would-be mixing of
the four-quark operators in class $B$, i.e.\ $Q^c_{7 \dots 10}$, with the QCD penguin operators and $Q_{9V}$.\footnote{The absence
of this mixing follows from the fact that the chiralities of the $(\bar s, b)$ pair of fields differ between these
groups of operators. Chirality conservation of massless QCD prevents mixing of operators of the same dimension but
distinct quark chiralities, even for nonvanishing quark masses. For the same reason, $Q^c_{5, 6}$ do not mix into $Q_{9V}$.}
 As a result
of the structure of (\ref{eq:adm}),  $\vec C_A$ and $\vec C_B$ do not mix; their renormalization, including the mixing of each
into the QCD penguins, dipoles, and $C_{9V}$, can therefore be considered independently. The evolution in the situation where only
one of $\vec C_A$ or $\vec C_B$ is present is described by an anomalous dimension of
the form \eqref{eq:admblock}, with $\hat \gamma_{cc} \to \gamma_{XX}$, $\hat \gamma_{cp} \to \hat \gamma_{Xp}$,
$\vec \gamma_{c8} \to \vec \gamma_{X8}$, $\vec \gamma_{c7} \to \vec \gamma_{X7}$,
and $\vec \gamma_{c9} \to \vec \gamma_{X9}$, where $X = A$ or $B$ as applicable; let us refer to these limiting cases as `case A' and
`case B', respectively.
\begin{table}[ht]
\begin{center}
\begin{tabular}{|l|c|c|}
\hline
Block& Loop Order&Coupling Order\\
\hline 
$\hat{\gamma}_{AA}, \hat{\gamma}_{BB}, \hat{\gamma}_{Ap},  \hat{\gamma}_{pp}$ &\multirow{2}{*}{1} & \multirow{2}{*}{$\alpha_s^{(1)}$} \\
$\gamma_{77}$,  $\gamma_{87}$,  $\gamma_{88}$, ${\vec{\gamma}_{p9}}$ &  &   \\
\hline
$\vec{\gamma}_{A7}$, $\vec{\gamma}_{A8}$, $\vec{\gamma}_{p7}$, $\vec{\gamma}_{p8}$&2 & $\alpha_s^{(1)}$ \\
\hline
$\vec{\gamma}_{B7}$,   $\vec{\gamma}_{B8}$, $\vec{\gamma}_{A9}$ &1 & $\alpha_s^{(0)}$ \\
\hline
\end{tabular}
\caption{}
\label{tab:looporders}
\end{center}
\end{table}

Let us first consider case A, which includes the subset of operators discussed in \cite{Jager:2017gal} and in particular
includes the SM case. Here, a rescaling
\begin{equation}
  Q_{9V}(\mu) = \frac{\alpha_s(\mu)}{4\pi} \tilde Q_{9V}(\mu),
  \qquad
  C_{9V}(\mu) = \frac{4\pi}{\alpha_s(\mu)} \tilde C_{9V}(\mu),
\end{equation}
is sufficient to bring the anomalous-dimension matrix into the form \eqref{eq:admlo}, such that the solution \eqref{eq:rgesollo} applies (note that with the rescaling, $\gamma_{99} \to - 2 \beta_0$). This also shows that $C_{9V}$ formally starts at order $1/\alpha_s$ when $\vec C_A$ is nonzero (which includes the SM case).
The blocks $\vec \gamma_{A7}$, $\vec \gamma_{A8}$, $\vec \gamma_{p7},$ and $\vec \gamma_{p8}$, due to two-loop diagrams,
are scheme-dependent and induce scheme dependence of $C_{7\gamma}$ and $C_{8g}$ already at the leading order.
The scheme-dependence ultimately cancels out in observables. It is convenient and customary to define effective
dipole coefficients \cite{Buras:1993xp,Greub:1996tg}
\begin{eqnarray}
   C_{7\gamma}^{\rm eff}(\mu) &=& C_{7\gamma}(\mu) + \vec y \cdot \vec C_{Ap},  \\[2mm]
   C_{8g}^{\rm eff}(\mu) &=&  C_{8g}(\mu) + \vec z \cdot \vec C_{Ap}  ,
\end{eqnarray}
in such a fashion that the leading-order expression for $\mathcal{B}(B \to X_s \gamma)$ is proportional to $|C_{7\gamma}^{\rm eff}|^2$
(or $|C_{7\gamma}^{\rm eff}|^2 + |C_{7\gamma}^{\prime \rm eff}|^2$, in the presence of primed operators). This ensures that
$C_{7\gamma}^{\rm eff}$ and $C_{8\gamma}^{\rm eff}$ are scheme-independent to leading order.
In the CBSM scenario, with our choice of operator basis and renormalization,
including in particular anticommuting $\gamma_5$, the vectors $\vec{y}$, $\vec{z}$ and $\vec C_{Ap}$ read
\begin{eqnarray}
 \vec{y}&=& \left(0,0,0,0, \frac{2m_c}{m_b}, \frac{2m_c}{3m_b}, -\frac{1}{3},-\frac{4}{9},-\frac{20}{3},-\frac{80}{3} \right),\\
 \vec{z}&=& \left( 0,0,0,0, 0, \frac{m_c}{m_b}, 1,-\frac{1}{6},20,-\frac{10}{3} \right), \\
 \vec C_{Ap} &=& \left( C_1, C_2, C_3, C_4, C_5, C_6, C^p_3, C^p_4, C^p_5, C^p_6 \right) ,
\end{eqnarray}
extending the SM case \cite{Chetyrkin:1996vx,Buras:2000if,Gambino:2003zm,Bobeth:2003at}.

As in the SM case, one can choose to go to a special scheme where the vectors $\vec y$ and $\vec z$ vanish by
means of a finite renormalization of the four-quark operators
$$
   Q^A_i \to Q^A_i - 
 y_i  Q_{7\gamma} - \
z_i  Q_{8g} ,
$$
where $Q^A_i$ is any of the four-quark operators whose Wilson coefficients
appear in $\vec C_{Ap}$. In this special scheme, $C_{7\gamma}$ and $C_{8g}$ are (to leading order) equal to
$C_{7\gamma}^{\rm eff}$ and $C_{8g}^{\rm eff}$, respectively, by construction. The mixing among the four-quark operators is unaffected
by the change of scheme, but the anomalous dimension elements involving the dipole operators change.
The resulting leading-order anomalous dimension matrix in the special
scheme is sometimes referred to as $\gamma^{(0), \rm eff}$.

Let us now turn to case $B$. In this case, the mixing of the CBSM operators into the dipoles through $\vec \gamma_{B7}$ and
$\vec \gamma_{B8}$ arises at one-loop order, but there is no mixing into the semileptonic coefficient $C_{9V}$.
To restore the form \eqref{eq:admlo} we should now rescale the electromagnetic and chromomagnetic dipoles as
\begin{eqnarray}
  Q_{7\gamma}(\mu) = \frac{\alpha_s(\mu)}{4\pi} \tilde Q_{7\gamma}(\mu),
  \qquad
  C_{7\gamma}(\mu) = \frac{4\pi}{\alpha_s(\mu)} \tilde C_{7\gamma}(\mu),
  \\
  Q_{8g}(\mu) = \frac{\alpha_s(\mu)}{4\pi} \tilde Q_{8g}(\mu),
  \qquad
  C_{8g}(\mu) = \frac{4\pi} {\alpha_s(\mu)}\tilde C_{8g}(\mu).
\end{eqnarray}
This ensures that $\vec \gamma_{B7,8}$ begin at ${\cal O}(\alpha_s)$; at the same time it makes the
blocks $\vec \gamma_{p7}$ and $\vec \gamma_{p8}$ vanish at  ${\cal O}(\alpha_s)$. As was the situation for the semileptonic coefficients
in case A, now the dipole coefficients are formally of order ${\cal O}(1/\alpha_s)$. Note that this makes the Standard Model contribution
formally subleading.
This is reflected in the fact that we will find below that $\mathcal{B}(B \to X_s \gamma)$
poses extremely stringent constraints on $\vec C_B$, corresponding to BSM scales of tens of TeV. 
The leading-order dipole coefficients are now scheme-independent and are not complemented
by finite leading-order contributions from the four-quark operators at leading order, i.e.\ there is no need to define effective dipole coefficients.

Since $\vec C_A$ and $\vec C_B$ do not mix under renormalization, and the mixing into the penguin and dipole coefficients is simply additive, there
is no problem
patching together the solutions for both cases
 into an unambiguous and scheme-independent evolution operator. The part of the evolution matrix relevant
for our phenomenology corresponds to
the subset of coefficients
$\vec{C}(\mu_b)=(\vec{C}^c(\mu_b),C^{\mathrm{eff}}_{7\gamma}(\mu_b),C_{9V}(\mu_b))$, and with initial conditions  $\vec{C}(\mu_0)=(\vec{C}^c(\mu_0),0,0)$
for $\mu_0 = M_W$ and $\mu_b = 4.2$ GeV this reads
\begin{equation}\label{eq:Utot}
\scalebox{0.95}{\(
\left( \begin{array}{c} 
C_1^c (\mu_b) \\ C_2^c(\mu_b) \\ C_3^c(\mu_b) \\ C_4^c(\mu_b) \\ C_5^c(\mu_b) \\ C_6^c(\mu_b)
\\ C_7^c(\mu_b) \\ C_8^c(\mu_b) \\ C_9^c(\mu_b) \\ C_{10}^c(\mu_b) \\ C_{7\gamma}^{\rm eff}(\mu_b) \\ C_{9V}(\mu_b)
\end{array} \right) \!=\!
\left(
\begin{array}{cccccccccc}
 1.1 & -0.27 & 0 & 0 & 0 & 0 & 0 & 0 & 0 & 0 \\
 -0.27 & 1.1 & 0 & 0 & 0 & 0 & 0 & 0 & 0 & 0 \\
 0 & 0 & 0.92 & 0 & 0 & 0 & 0 & 0 & 0 & 0 \\
 0 & 0 & 0.33 & 1.9 & 0 & 0 & 0 & 0 & 0 & 0 \\
 0 & 0 & 0 & 0 & 1.9 & 0.33 & 0 & 0 & 0 & 0 \\
 0 & 0 & 0 & 0 & 0 & 0.92 & 0 & 0 & 0 & 0 \\
 0 & 0 & 0 & 0 & 0 & 0 & 1.0 & 0.05 & 2.70 & 1.70 \\
 0 & 0 & 0 & 0 & 0 & 0 & 0.37 & 2.0 & 2.30 & -0.55 \\
 0 & 0 & 0 & 0 & 0 & 0 & 0.07 & 0.07 & 1.80 & 0.04 \\
 0 & 0 & 0 & 0 & 0 & 0 & 0.01 & -0.02 & -0.29 & 0.82 \\
 0.02 & -0.19 & -0.015 & -0.13 & 0.56 & 0.17 & -1.0 & -0.47 & 4.00 & 0.70 \\
 8.50 & 2.10 & -4.30 & -2.00 & 0 & 0 & 0 & 0 & 0 & 0 \\
\end{array}
\right)\!
\left( \begin{array}{c} 
C_1^c (M_W) \\ C_2^c(M_W) \\ C_3^c(M_W) \\ C_4^c(M_W) \\ C_5^c(M_W) \\ C_6^c(M_W)
\\ C_7^c(M_W) \\ C_8^c(M_W) \\ C_9^c(M_W) \\ C_{10}^c(M_W) 
\end{array} \right)
\)}
.\end{equation}

\subsubsection{Remarks on computed and uncomputed ADM elements}
\label{sub:remarks_adm}

As part of our earlier work \cite{Jager:2017gal}, we made the first calculation of the mixing of the BSM operators \(Q_{3,4}^c\) into the operators \(P_{3-6}\), \(Q_{7\gamma}\), and \(Q_{9V}\).
The mixing into the photon penguin \(Q_{7\gamma}\) is the most technically challenging, arising from two-loop diagrams.
We did not compute the two-loop mixing of \(Q_{3,4}^c\) into the gluon penguin \(Q_{8g}\), and neglect this mixing in our numerical results in both the previous work and this article -- the corresponding result for the mixing of the SM operators \(Q^c_{1,2}\) is known to produce only a very small effect on the contribution of \(C_2 (M_W)\) to \(C_{7\gamma}^\text{eff} (\mu_b)\), and we expect a similarly small effect from the results we neglect.
Further details can be found in the Appendix of \cite{Jager:2017gal}.
\section{Observables} 
\label{sec:observables}
In this section we collect a set of observables that are very sensitive to $(\bar{c} b) (\bar{s} c)$ operators and 
allow us thus to constrain the possible size of new $b \to c \bar{c} s$ contributions:
the dominant weak annihilation contribution to the $B_s$ lifetime, $\tau (B_s)$, is given by a $(\bar{c} b) (\bar{s} c)$ transition,
which is also the leading term to the mixing induced decay rate difference of neutral $B_s$ mesons, $\Delta \Gamma_s$.
Taking a $(\bar{c} b) (\bar{s} c)$ operator and closing the charm quarks to a loop we get large penguin contributions,
that sizeably affect $b \to s \gamma$ and $ b \to s \ell \ell$ decays. Finally the gold-plated mode for CP-violation, $B_d \to J/\psi K_S$
is triggered by a tree-level $b \to c \bar{c} s$ decay.
Below we determine the dependence of all these observables on the new four quark operators.
 \subsection{Lifetime ratio \texorpdfstring{$\tau (B_s) / \tau (B_d)$}{tau(Bs)/tau(Bd)}}
 \label{sub:lifetime}
 The total decay width of the $B_s$ meson can be expressed as
\begin{equation}\label{eq:Gammas}
              \Gamma_s\equiv \frac{1}{\tau(B_s)} = \frac{1}{2M_{B_s}}\langle B_s\mid\mathcal{T}\mid B_s \rangle \, ,
\end{equation}
with the transition operator
\begin{equation}\label{eq:transition1}
\mathcal{T}=\mathrm{Im}\left[ i\int d^4xT[\mathcal{H}_{\mathrm{eff}}(x)\mathcal{H}_{\mathrm{eff}}(0)]\right]\\.
\end{equation}
According to the Heavy Quark Expansion (HQE) (see \cite{Lenz:2014jha} for a review and early references)
the transition operator can be expanded in inverse powers of the
heavy $b$ quark mass -- each term in the expansion contains perturbative Wilson coefficients and non-perturbative
matrix elements of $\Delta B = 0$ operators.
We will investigate the precisely measured lifetime ratio
\begin{eqnarray}
  \frac{\tau (B_s)}{\tau (B_d) } & = &  1 + \frac{\Gamma_d - \Gamma_s}{\Gamma_s}
  \approx 1 + \frac{\Gamma_d^{\rm SM} - \Gamma_s^{\rm SM}}{\Gamma_s^\text{SM}} - \frac{\Gamma_s^{\rm BSM}}{\Gamma_s}
  =  \frac{\tau (B_s)^{\rm SM}}{\tau (B_d)^{\rm SM} } - \Gamma_s^{\rm BSM} \tau (B_s)^{\rm exp} \, .
\label{eq:tBstBd}
\end{eqnarray}
The SM value of $\tau (B_s) / \tau (B_d) $ will be taken from \cite{Kirk:2017juj}, which uses
perturbative input from \cite{Beneke:2002rj,Franco:2002fc}.
The leading contribution to this lifetime ratio is given by the third order in the HQE (see e.g. \cite{Lenz:2014jha})
and the dominant contribution at this order to the $B_s$ decay rate is given by $\Delta B = 1 $ $b \to c \bar{c} s$
transitions. These transitions are CKM suppressed for the $B_d$ meson.
Thus we assumed in \eqref{eq:tBstBd}  that the BSM effects due to the new  $b \to c \bar{c} s$ operators
contribute only to $\Gamma_s^{\rm BSM}$ and the $B_d$ lifetime agrees with the SM expectation.
The leading (i.e. LO-QCD and dimension six in the HQE) new $b \to c \bar{c} s$ contributions are given by
\begin{align}
\Gamma_s^{\rm BSM} & = \frac{8G_F^2}{M_{B_s}} \left| \lambda_c \right|^2 \sum\limits_{i=1}^{20}\sum\limits_{j=1}^{20} C^c_i(C^c_j)^* \langle B_s\mid \mathrm{Im}\left\{i \int d^4xT\left[Q^c_{i}(x)Q_{j}^{ c\dagger}(0)\right]\right\}\mid B_s \rangle - \Gamma_s^{\rm SM} \nonumber
\\
&= \frac{G_F^2 m_b^2 M_{B_s} f_{B_s}^2}{4 \pi} N_c \sqrt{1-z} \left| \lambda_c \right|^2
  \left[ \sum\limits_{i=1}^{20}\sum\limits_{j=1}^{20} C_i^c(C_j^c)^* \Gamma(i,j)
        -\sum\limits_{i=1}^{2}\sum\limits_{j=1}^{2} C^{c, \rm SM}_i(C^{c,\rm SM}_j)^* \Gamma(i,j)
        \right]  \, . \label{eq:gamcalc}
\end{align}
The ratio of charm and bottom quark mass is denoted by $z=4m_c^2 / m_b^2$. 
To avoid a double counting of the SM contribution due to $Q_{1}^c$ and $Q_2^c$, we subtract explicitly the SM contributions proportional to
$C^{c,\rm SM}_i(C^{c,\rm SM}_j)^*$.
The terms $ \Gamma(i,j)$ are symmetric ($ \Gamma(i,j) =  \Gamma(j,i)$) and they can be further split up into contributions
of eight different $\Delta B = 0$ four-quark operators.
\begin{eqnarray}
  Q_{XY} & = & \left( \bar{b} \gamma_\mu P_X s \right)  \left(\bar{s} \gamma^\mu P_Y b \right) \, ,
  \qquad \qquad
  Q^S_{XY}  = \left( \bar{b}  P_X s \right)  \left(\bar{s}  P_Y b \right) \, ,
  \nonumber \\
  \tilde{Q}_{XY} & = & \left( \bar{b}^i \gamma_\mu P_X s^j \right)  \left(\bar{s}^j \gamma^\mu P_Y b^i \right) \, ,
  \qquad
  \tilde{Q}^S_{XY}   = \left( \bar{b}^i            P_X s^j \right) \left(\bar{s}^j            P_Y b^i \right) \, ,
\end{eqnarray}
with $XY = LL, LR$. The matrix elements of these operators are parameterised as
\begin{equation}
\begin{aligned}
  \frac{1}{2M_{B_s}} \langle B_s | {\mathop{Q}\limits^{\scriptscriptstyle(\sim)}}_{\!LL}, {\mathop{Q}\limits^{\scriptscriptstyle(\sim)}}_{\!LR} | {B}_s \rangle
  &= \frac{1}{8} {\mathop{B}\limits^{\scriptscriptstyle(\sim)}}_{\!1,3} (\mu) f_{B_s}^2 M_{B_s} \, , \\
  \frac{1}{2M_{B_s}} \langle B_s | {\mathop{Q}\limits^{\scriptscriptstyle(\sim)}}{}^{\!S}_{\!LR}, {\mathop{Q}\limits^{\scriptscriptstyle(\sim)}}{}^{\!S}_{\!LL} | {B}_s \rangle
  &= \frac{1}{8} \frac{M_{B_s}^2}{[\bar{m}_{b}(\bar{m}_{b})+\bar{m}_{s}(\bar{m}_{b})]^{2}} {\mathop{B}\limits^{\scriptscriptstyle(\sim)}}_{\!2,4} (\mu) f_{B_s}^2 M_{B_s} \,,
\end{aligned}
\end{equation}
with the decay constant $f_{B_s}$. The bag parameters ${\mathop{B}\limits^{\scriptscriptstyle(\sim)}}_{\!1,2}$ have been
determined in \cite{Kirk:2017juj}, while for the remaining bag parameters we will use vacuum insertion approximation:
\begin{equation}
B_{3,4} =- 1 \,, \qquad \widetilde{B}_{3,4} = - \frac{1}{N_c} \,.
\end{equation}
For the individual contributions we get (where we define
\begin{equation}
{\mathop{B}\limits^{\scriptscriptstyle(\sim)}}{}^{\prime}_{\!2,4} = \frac{M_{B_{s}}^{2}}{[\bar{m}_{b}(\bar{m}_{b})+\bar{m}_{s}(\bar{m}_{b})]^{2}} {\mathop{B}\limits^{\scriptscriptstyle(\sim)}}_{\!2,4} 
\end{equation}
for the sake of brevity)
\begin{align}
\Gamma(1,1) & = \frac{1}{12}  \left[ 2 (z+2) B'_2+ (z-4) B_1 \right] \,, &\Gamma(1,3) &= \frac{1}{8 }  z B_1  \,, &\Gamma(1,5)  &=  - \frac{1}{2 } \sqrt{ z}  B'_2  \, , \nonumber \\
\Gamma (1,9)  &=  \frac{1}{2 } \sqrt{z} (4 B'_2-B_1 ) \, , &\Gamma (1,11) & =  -\frac{1}{4 } z B_3 \, , &\Gamma (1,7) & = \frac{1}{8 } \sqrt{z} B_1 \, , \nonumber \\
\Gamma (1,13) & =  -\frac{1}{24 } \left[ 2 (z+2) B'_4 +(z-4) B_3 \right] \, , &\Gamma (1,15) & =  \frac{1}{2 } \sqrt{z} B'_4  \, , \\
\Gamma (1,17) & =  \frac{1}{8 }\sqrt{z} \left[ B_3 -2  B'_4 \right] \, , &\Gamma (1,19)  &=  -\frac{1}{2 }\sqrt{z} \left[ 2  B'_4 + B_3 \right]  \,. \nonumber \hspace*{-5em}
\end{align}

\begin{equation}
\begin{aligned}
\Gamma (3,3) &= \frac{1}{4} \Gamma (1,1) \,, &\Gamma (3,7) &=   \frac{1}{16 } \sqrt{z}  \left(2 B'_2 -B_1 \right)  \, , &\Gamma (3,5) & = \frac{1}{2} \Gamma(1,5) \, , \\
\Gamma(3,13) &= -\frac{1}{16 } zB_3 \, , &\Gamma (3,9) &= \frac{1}{4 } \sqrt{z}  \left( B_1 + 2 B'_2 \right) \, , &\Gamma (3,15) & = \frac{1}{2} \Gamma(1, 15) \, ,  \\
\Gamma(3,17) &=  -\frac{1}{16 } \sqrt{z} B_3 \, , &\Gamma (3,19) & =\frac{1}{4 } \sqrt{z} \left( B_3  - 4 B'_4 \right) \,.
\end{aligned} 
\end{equation}

\begin{equation}
\begin{aligned}
\Gamma(5,5) & = (2-z) B'_2 \,, &\Gamma(5,7) &= - \frac{1}{4 } z B'_2 \,, &\Gamma(5,9)  &= 12 \Gamma(5,7)\, ,\\
\Gamma(5,15) &= -z B'_4 \, , &\Gamma(5,17) &= \frac{1}{4 } (2-z) B'_4 \,, &\Gamma (5,19) &= 12 \Gamma(5,17) \,.
\end{aligned}
\end{equation}

\begin{equation}
\begin{aligned}
\Gamma (7,7) & = -\frac{1}{48 } \left[ (z+2)B_1 + 2 (z-4) B'_2 \right] \, , &\Gamma (7,17) &= \frac{1}{4} \Gamma(1,11) \, ,  \\
\Gamma (7,9) & = \frac{1}{12 } \left[ (z+2)B_1 + 2 (8-5z) B'_2 \right] \, ,  &\Gamma(7,19) & = \frac{1}{4 } z \left( B_3 - 4 B'_4 \right) \,.
\end{aligned}
\end{equation}

\begin{equation}
\begin{aligned}
\Gamma (9,9) & = -\frac{1}{3 } \left[ (z+2) B_1 + 2 (13z-28) B'_2 \right] \, , &\Gamma (9,19) &= -z \left( 8 B'_4 + B_3 \right)\,.
\end{aligned}
\end{equation}
All remaining terms can be extracted from those given via the following rules:
\begin{equation}
\Gamma (i,j) = 
\begin{cases}
  \left. \frac{1}{N_c} \Gamma(i-1, j-1) \right|_{B^{(')}_i \to \tilde{B}^{(')}_i} &\text{ for $i,j$ even,} \\
  \frac{1}{N_c} \Gamma(i-1, j) &\text{ for $i$ even, $j$ odd,} \\
  \frac{1}{N_c} \Gamma(i, j-1) &\text{ for $i$ odd, $j$ even,} \\
  \Gamma(i-10,j-10)  &\text{ for $i,j > 10$,}  \\
  \Gamma(j-10, i+10) &\text{ for $i<10, j>10, i> j-10$,} \\
  \Gamma(j,i) & \text{ for $i > j$.}
\end{cases} \label{eq:gammaijrules}
\end{equation}
The interested reader can download a \texttt{Mathematica} program containing the full algebraic expressions from the arXiv version of this article .
 \subsection{\texorpdfstring{$B_s$}{Bs} mixing observables \texorpdfstring{$\Delta \Gamma_s$}{Delta Gamma s} and \texorpdfstring{$a_{sl}^s$}{asls}}
 \label{sub:Bmix}
 The decay rate difference of neutral $B_s$ mesons, $\Delta \Gamma_s$, and the flavour specific CP asymmetry in $B_s$ decays, $a_{sl}^s$,
 are sensitive to new $ b \to c \bar{c} s$ effects.
 Using the conventions of  \cite{Artuso:2015swg} we get
 \begin{eqnarray}
  \Delta\Gamma_s  = 2|\Gamma_{12}^s|\cos\phi_{12}^s \, ,
  &\hspace{1cm} &
  a^s_{sl} = \left|\frac{\Gamma_{12}^s}{M_{12}^s}\right|\sin\phi_{12}^s \, ,
 \end{eqnarray}
 where $\Gamma_{12}^s$ denotes the absorptive part of the mixing diagrams and  $M_{12}^s$ the dispersive part.
 The relative phase of these contributions is defined as
\begin{equation}
\phi_{12}^s:=\mathrm{arg}\left(-\frac{M^s_{12}}{\Gamma^s_{12}}\right)\\.
\end{equation}
Similar to the case of the total decay rate the off diagonal matrix element of the absorptive part of the
$B_s-\bar{B}_s$ mixing matrix, $\Gamma^s_{12}$ can be expressed in terms of double insertion of the effective Hamiltonian
 \begin{eqnarray}\label{eq:gam12}
  \Gamma_{12}^s  =  \frac{1}{2M_{B_s}}\langle B_s| \mathcal{T}| \bar{B}_s\rangle \, ,
  & \hspace{0.5cm}  \mbox{with} \hspace{0.5cm} &
  \mathcal{T} =  \mathrm{Im}\left[ i\int d^4xT[\mathcal{H}_{\mathrm{eff}}(x)\mathcal{H}_{\mathrm{eff}}(0)]\right]\, .
\end{eqnarray}
This matrix element can be split up into a SM contribution and a BSM contribution due to the new $b \to c \bar{c} s$ transitions
\begin{eqnarray}
  \Gamma_{12}^s & = & \Gamma_{12}^\text{SM}+\Gamma_{12}^{c\bar{c}} \, .
 \end{eqnarray}
The numerical value of the SM part is based on \cite{Beneke:1998sy,Beneke:2003az,Ciuchini:2003ww,Lenz:2006hd}, the
BSM part is further decomposed as
 \begin{align}
\Gamma_{12}^{c\bar{c}} &=
  \frac{4G_F^2 \lambda_c^2}{M_{B_s}}\sum\limits_{i=1}^{20}\sum\limits_{j=1}^{20}C_i^cC_j^c \langle B_s|\textrm{Im}\left\{i\int d^4 xT[Q^c_i(x)Q_j^c(0)]\right\}| \bar{B}_s\rangle \nonumber
\\
&= \frac{G_F^2\lambda_c^2 m_b^2M_{B_s}f_{B_s}^{2}}{12 \pi} \sqrt{1-z} \left[8 G(z) B +F(z) \tilde{B}^{\prime}_S + \frac{1}{4} H(z) B'_{4} + \frac{1}{12} J(z) B'_{5} \right]
\\
&\qquad +\mathcal{O}\left(\frac{\Lambda_{QCD}}{m_b}\right)+\mathcal{O}(\alpha_s) \,.
\label{eq:gamma12bsm}
\end{align}
The arising four quark operators can be parametrised as
\begin{equation}
\begin{aligned}
\langle Q\rangle &= \frac{8}{3}M_{B_s}^{2}f_{B_s}^{2} B \, ,
&\quad
\langle \tilde{Q}_{S}\rangle &= \frac{1}{3}M_{B_s}^{2}f_{B_s}^{2} \tilde{B}^{\prime}_S \, ,
\\
\langle {Q}_{4}\rangle &= \frac{1}{2} M_{B_s}^{2}f_{B_s}^{2} B'_{4} \, ,
&\quad
\langle {Q}_{5}\rangle &= \frac{1}{6} M_{B_s}^{2}f_{B_s}^{2} B'_{5} \, ,
\end{aligned}
\label{MixME}
\end{equation}
with
\begin{equation}
\begin{aligned}
\tilde{B}_{S}^{\prime}&=\frac{M_{B_{s}}^{2}}{(\bar{m}_{b}(\bar{m}_{b})+\bar{m}_{s}(\bar{m}_{b}))^{2}}\tilde{B}_{S} \,,
\\
B_{4}^{\prime}&=\left( \frac{M_{B_{s}}^{2}}{(\bar{m}_{b}(\bar{m}_{b})+\bar{m}_{s}(\bar{m}_{b}))^{2}} + \frac{1}{6} \right) {B}_{4} \,,
\\
B_{5}^{\prime}&=\left( \frac{M_{B_{s}}^{2}}{(\bar{m}_{b}(\bar{m}_{b})+\bar{m}_{s}(\bar{m}_{b}))^{2}} + \frac{3}{2} \right) {B}_{5} \,,
\end{aligned}
\end{equation}
which matches the definitions in \cite{Artuso:2015swg} for $Q$ and $\tilde{Q}_s$ and \cite{Bazavov:2016nty,King:2019lal,Dowdall:2019bea} for $Q_4$ and $Q_5$; these bag parameters have been recently determined in
\cite{Bazavov:2016nty,King:2019lal,Dowdall:2019bea}.
The coefficient functions read
\allowdisplaybreaks[1]
\begin{eqnarray}
  F(z) &=& \left( 1 + \frac{z}{2} \right) \! \!\left[ \frac{\cij{1}{1} - (C_1^{c, \rm SM})^2}{2} + \frac{ \cij{1}{2} - C_1^{c,\rm SM} C_2^{c,\rm SM}  }{3} - \frac{ \cij{2}{2} - (C_2^{c,\rm SM})^2 }{6} + \frac{ \cij{3}{3} }{8}+ \frac{\cij{3}{4} }{12} - \frac{ \cij{4}{4} }{24} \right] 
\nonumber \\ && - \left(1 - \frac{z}{2} \right)  \left[  
 18 \, \cij{5}{9} + 6 ( \cij{5}{10}  +  \cij{6}{9} - \cij{6}{10} ) + \frac{3}{2} \cij{5}{7} 
+ \frac{ \cij{5}{8} + \cij{6}{7}- \cij{6}{8} }{2}
\right] 
\nonumber \\
&& + \, \sqrt{z} \left[  6 \cij{1}{9} + 2\cij{1}{10} + 2\cij{2}{9} - 2\cij{2}{10} - \frac{3}{2} ( \cij{1}{5} - \cij{3}{9} )
- \frac{3}{4} \cij{3}{5}  + \frac{3}{8} \cij{3}{7}
    \right. \nonumber \\ && \qquad \left. 
        - \frac{\cij{1}{6} + \cij{2}{5} - \cij{2}{6} - \cij{3}{10} - \cij{4}{9} + \cij{4}{10} }{2}
- \frac{ \cij{3}{6} + \cij{4}{5} - \cij{4}{6}  }{4}
     \right. \nonumber \\ && \qquad \left.
+ \frac{ \cij{3}{8} \!+\! \cij{4}{7} \!-\! \cij{4}{8}   }{8}
\right] 
   \! + \! z \! \left[ 15 \cij{9}{9} + 10 \cij{9}{10} - 5 \cij{10}{10} \!+ \frac{3}{2} \cij{7}{9} + \! \frac{3}{2} \cij{5}{5} + \! \cij{5}{6}
     \right. \nonumber \\ && \qquad  \left. 
 + \frac{ \cij{7}{10} + \cij{8}{9} - \cij{8}{10} }{2}
                                           - \frac{ \cij{6}{6}}{2} + \frac{\cij{7}{8}   }{8} + \frac{ 3 \cij{7}{7} -\cij{8}{8} }{16}                                        
\right]  \nonumber \\
&& + \left( C_i ^c C_j^c  \to C^{c'}_i C^{c'}_j \, , \, C_{1,2}^{c,\text{SM}} \to 0 \right) \,,
\end{eqnarray}
\begin{eqnarray}
  G(z) &=& - \left(1 - \frac{z}{2} \right) \left[ 9 \cij{5}{9} + 3 ( \cij{5}{10} + \cij{6}{9} ) + \frac{3 \cij{5}{7} + \cij{5}{8} + \cij{6}{7} }{4} \right]
\nonumber \\ && 
- \left( 1 - z \right) \left[ \frac{\cij{1}{1} - (C_1^{c, \rm SM})^2 }{4} + \frac{\cij{1}{2} - C_1^{c, \rm SM} C_2^{c, \rm SM}}{6}+ \frac{ \cij{3}{3} }{16} + \frac{ \cij{3}{4} }{24} \right] \nonumber \\ &&
 - \left( 1 - \frac{z}{4} \right) \left[ \frac{ \cij{2}{2} - (C_2^{c, \rm SM})^2}{6} + \frac{ \cij{4}{4} }{24} \right] + z \left[ 6 \cij{9}{9} + 4 \cij{9}{10} + \frac{3}{2} \cij{7}{9}  -\frac{\cij{8}{8}}{32} \right. \nonumber \\ &&
 \left. \qquad + \frac{ \cij{7}{10} + \cij{8}{9} - \cij{10}{10}  }{2}  + \frac{ \cij{8}{10} }{4}
             + \frac{3}{4} \cij{5}{5} + \frac{ \cij{5}{6}  }{2}  \right. \nonumber \\ &&
\left. \qquad + \frac{ 3 \cij{1}{3} + \cij{1}{4} + \cij{2}{3} + \cij{2}{4} }{8}  \right]
+ \sqrt{z} \left[ \frac{3}{2} ( \cij{1}{9}  + \cij{3}{9} ) - \frac{3}{4} \cij{1}{5} \right. \nonumber \\ && 
\qquad  + \frac{ \cij{1}{10} + \cij{2}{9} - \cij{2}{10} + \cij{3}{10} +  \cij{4}{9} }{2} +  \frac{3}{8} ( \cij{1}{7} - \cij{3}{5} )   \nonumber \\ && 
\qquad \left. - \frac{ \cij{1}{6} + \cij{2}{5}   - \cij{4}{10} }{4} + \frac{ \cij{1}{8} + \cij{2}{7} + \cij{2}{8} - \cij{3}{6} - \cij{4}{5}}{8} 
        - \frac{ \cij{4}{8} }{16}
       \right] \nonumber \\ && 
       + \left( C_i^c C_j^c \to C_i^{c'} C_j^{c'} \, , \, C_{1,2}^{c,\text{SM}} \to 0 \right) \,, \label{eq:gamma12G}
\end{eqnarray}
\begin{align}
H(z) =
z
&\left[
2 (3 C^c_3+C^c_4) C^{c'}_{1}+2 (6 C^c_2+C^c_3+C^c_4) C^{c'}_{2}+3 C^c_4 C^{c'}_{4} +2 C^c_2 (C^{c'}_{3}+C^{c'}_{4})-24 C^c_5 C^{c'}_{6}
\right.
\nonumber \\
&\left.
+2 C^c_1 (3 C^{c'}_{3}+C^{c'}_{4})-6 (12 C^c_5+4 C^c_6+3 C^c_7+C^c_8+36 C^c_9) C^{c'}_{5}-6 C^c_7 C^{c'}_{6}-18 C^c_5 C^{c'}_{7}
\right.
\nonumber \\
&\left.
-6 C^c_6 C^{c'}_{7}-6 C^c_7 C^{c'}_{7}-2 C^c_8 C^{c'}_{7}-6 C^c_5 C^{c'}_{8}-2 C^c_7 C^{c'}_{8}+C^c_8 C^{c'}_{8} -72( 3C^c_5-C^c_6) C^{c'}_{9}
\right.
\nonumber \\
&\left.
-48 C^c_7 C^{c'}_{9}-16 C^c_8 C^{c'}_{9}-4 C^c_8 C^{c'}_{10}-4 C^c_{10} (18 C^{c'}_{5}+4 C^{c'}_{7}+C^{c'}_{8}+56 C^{c'}_{9}-4 C^{c'}_{10})
\right.
\nonumber \\
&\left.
-72 C^c_5 C^{c'}_{10}-16 C^c_7 C^{c'}_{10}-8 C^c_9 (9 C^{c'}_{6}+6 C^{c'}_{7}+2 C^{c'}_{8}+84 C^{c'}_{9}+28 C^{c'}_{10})
\right]
\nonumber \\
+ 3 \sqrt{z}
&\left[
2 (-6 C^c_5-2 C^c_6+3 C^c_7+C^c_8+12 C^c_9) C^{c'}_{1}-4 C^c_5 C^{c'}_{2}+2 C^c_7 C^{c'}_{2}-2 C^c_8 C^{c'}_{2}-6 C^c_5 C^{c'}_{3}
\right.
\nonumber \\
&\left.
-2 C^c_6 C^{c'}_{3}+8 C^c_{10} (C^{c'}_{1}+C^{c'}_{2}+C^{c'}_{3})-2 C^c_5 C^{c'}_{4}+C^c_8 C^{c'}_{4}-4 C^c_{10} C^{c'}_{4}
\right.
\nonumber \\
&\left.
+8 C^c_9 (C^{c'}_{2}+3 C^{c'}_{3}+C^{c'}_{4})-2 (2 C^c_2+3 C^c_3+C^c_4) C^{c'}_{5}-2 C^c_3 C^{c'}_{6}+2 C^c_2 C^{c'}_{7}-2 C^c_2 C^{c'}_{8}
\right.
\nonumber \\
&\left.
+C^c_4 C^{c'}_{8}+8 C^c_2 C^{c'}_{9}+24 C^c_3 C^{c'}_{9}+8 C^c_4 C^{c'}_{9}+8 C^c_2 C^{c'}_{10}+8 C^c_3 C^{c'}_{10}-4 C^c_4 C^{c'}_{10}
\right.
\nonumber \\
&\left.
+2 C^c_1 (-6 C^{c'}_{5}-2 C^{c'}_{6}+3 C^{c'}_{7}+C^{c'}_{8}+12 C^{c'}_{9}+4 C^{c'}_{10})
\right]
\nonumber \\
+2
&\left[
2 C^c_1 (3 C^{c'}_{3}+C^{c'}_{4})+2 (3 C^c_3 C^{c'}_{1}+C^c_4 C^{c'}_{1}+C^c_3 C^{c'}_{2}-2 C^c_4 C^{c'}_{2}+C^c_2 C^{c'}_{3}-2 C^c_2 C^{c'}_{4})
\right.
\nonumber \\
&\left.
+24 (3 C^c_5 C^{c'}_{5}+C^c_6 C^{c'}_{5}+C^c_5 C^{c'}_{6})+3 C^c_7 C^{c'}_{7}+C^c_8 C^{c'}_{7}+C^c_7 C^{c'}_{8}+C^c_8 C^{c'}_{8}+60 C^c_7 C^{c'}_{9}
\right.
\nonumber \\
&\left.
+20 C^c_8 C^{c'}_{9}+20 C^c_7 C^{c'}_{10}-4 C^c_8 C^{c'}_{10}+4 C^c_{10} (5 C^{c'}_{7}-C^{c'}_{8}+52 C^{c'}_{9}+4 C^{c'}_{10})
\right.
\nonumber \\
&\left.
+4 C^c_9 (15 C^{c'}_{7}+5 C^{c'}_{8}+52 (3 C^{c'}_{9}+C^{c'}_{10}))
\right]
\,,
\end{align}
and
\begin{align}
J(z) =
z
&\left[
12 C^c_2 C^{c'}_{1}+6 C^c_3 C^{c'}_{1}+2 C^c_4 C^{c'}_{1}+2 C^c_3 C^{c'}_{2}+2 C^c_4 C^{c'}_{2}+2 C^c_2 C^{c'}_{3}+9 C^c_3 C^{c'}_{3}+3 C^c_4 C^{c'}_{3}
\right.
\nonumber \\
&\left.
+2 C^c_2 C^{c'}_{4}+3 C^c_3 C^{c'}_{4}+2 C^c_1 (18 C^{c'}_{1}+6 C^{c'}_{2}+3 C^{c'}_{3}+C^{c'}_{4})-24 C^c_6 C^{c'}_{6}-6 C^c_8 C^{c'}_{6}
\right.
\nonumber \\
&\left.
+3 C^c_7 C^{c'}_{7}+C^c_8 C^{c'}_{7}-12 C^c_9 C^{c'}_{7}-6 C^c_6 C^{c'}_{8}+C^c_7 C^{c'}_{8}-2 C^c_8 C^{c'}_{8}-4 C^c_9 C^{c'}_{8}-12 C^c_7 C^{c'}_{9}
\right.
\nonumber \\
&\left.
-4 C^c_8 C^{c'}_{9}+48 C^c_9 C^{c'}_{9}-72 C^c_6 C^{c'}_{10}-4 C^c_7 C^{c'}_{10}-16 C^c_8 C^{c'}_{10}+16 C^c_9 C^{c'}_{10}
\right.
\nonumber \\
&\left.
-4 C^c_{10} (18 C^{c'}_{6}+C^{c'}_{7}+4 C^{c'}_{8}-4 C^{c'}_{9}+56 C^{c'}_{10})
\right]
\nonumber \\
+3 \sqrt{z}
&\left[
-6 C^c_7 C^{c'}_{1}-2 C^c_8 C^{c'}_{1}+24 C^c_9 C^{c'}_{1}-4 C^c_6 C^{c'}_{2}-2 C^c_7 C^{c'}_{2}+2 C^c_8 C^{c'}_{2}+8 C^c_9 C^{c'}_{2}
\right.
\nonumber \\
&\left.
+3 C^c_7 C^{c'}_{3}+C^c_8 C^{c'}_{3}-12 C^c_9 C^{c'}_{3}-2 C^c_6 C^{c'}_{4}+C^c_7 C^{c'}_{4}-4 C^c_9 C^{c'}_{4}
\right.
\nonumber \\
&\left.
+4C^c_{10} (2 C^{c'}_{1}+2 C^{c'}_{2}- C^{c'}_{3}+2 C^{c'}_{4})-4 C^c_2 C^{c'}_{6}-2 C^c_4 C^{c'}_{6}-2 C^c_2 C^{c'}_{7}+3 C^c_3 C^{c'}_{7}
\right.
\nonumber \\
&\left.
+C^c_4 C^{c'}_{7} +2 C^c_2 C^{c'}_{8}+C^c_3 C^{c'}_{8}+8 C^c_2 C^{c'}_{9}-12 C^c_3 C^{c'}_{9}-4 C^c_4 C^{c'}_{9}+8 C^c_2 C^{c'}_{10}-4 C^c_3 C^{c'}_{10}
\right.
\nonumber \\
&\left.
+8 C^c_4 C^{c'}_{10} -2 C^c_1 (3 C^{c'}_{7}+C^{c'}_{8}-4 (3 C^{c'}_{9}+C^{c'}_{10}))
\right]
\nonumber \\
+2
&\left[
-12 C^c_3 C^{c'}_{1}-4 C^c_4 C^{c'}_{1}-4 C^c_3 C^{c'}_{2}+2 C^c_4 C^{c'}_{2}-4 C^c_2 C^{c'}_{3}+2 C^c_2 C^{c'}_{4}-4 C^c_1 (3 C^{c'}_{3}+C^{c'}_{4})
\right.
\nonumber \\
&\left.
+24 C^c_6 C^{c'}_{6}+3 C^c_7 C^{c'}_{7}+C^c_8 C^{c'}_{7}-12 C^c_9 C^{c'}_{7}+C^c_7 C^{c'}_{8}+C^c_8 C^{c'}_{8}-4 C^c_9 C^{c'}_{8}-12 C^c_7 C^{c'}_{9}
\right.
\nonumber \\
&\left.
-4 C^c_8 C^{c'}_{9}+48 C^c_9 C^{c'}_{9}-4 C^c_{10} (C^{c'}_{7}-5 C^{c'}_{8}-4 C^{c'}_{9}-52 C^{c'}_{10})-4 C^c_7 C^{c'}_{10}+20 C^c_8 C^{c'}_{10}
\right.
\nonumber \\
&\left.
+16 C^c_9 C^{c'}_{10}
\right]
\,.
\end{align}
The interested reader can download a \texttt{Mathematica} program containing the full algebraic expressions from the arXiv version of
this article.

 At this point we would like to mention that neglecting the CKM structure $\lambda_u$ (as advocated in Section \ref{sec:setup})
 might result in misleading conclusions for the semileptonic CP asymmetry -- depending on the experimental precision of the
 corresponding measurement.
 The mixing observables can also be determined via the following relations
 \begin{eqnarray}
   \frac{\Delta \Gamma_s}{\Delta M_s}  =   - {\cal R} \left( \frac{\Gamma_{12}^s}{M_{12}^s} \right) \, ,
   & \hspace{1cm} &
   a_{sl}^s  =  {\cal I} \left( \frac{\Gamma_{12}^s}{M_{12}^s}\right) \, .
 \end{eqnarray}
 Using the notation in \cite{Artuso:2015swg} and CKM unitarity ($\lambda_u + \lambda_c + \lambda_t =0$)
 we can further simplify
 \begin{eqnarray}
      - \frac{\Gamma_{12}^s}{M_{12}^s}  & = & \frac{     \lambda_c^2         \Gamma_{12}^{s,cc}
                         + 2 \lambda_c \lambda_u \Gamma_{12}^{s,uc}
                         +   \lambda_u^2         \Gamma_{12}^{s,uu}}{\lambda_t^2 \tilde{M}_{12}^s}
      \nonumber \\
       & = &  \frac{\Gamma_{12}^{s,cc}}{\tilde{M}_{12}^s}
        + 2 \frac{\lambda_u}{\lambda_t}
          \frac{\Gamma_{12}^{s,cc} -\Gamma_{12}^{s,uc}}{\tilde{M}_{12}^s}
        + \left(\frac{\lambda_u}{\lambda_t}\right)^2
          \frac{\Gamma_{12}^{s,cc} -2 \Gamma_{12}^{s,uc}+\Gamma_{12}^{s,uu} }{\tilde{M}_{12}^s}
      \nonumber
      \\
      & = & - 10^{-4}
       \left[ c + a \frac{\lambda_u}{\lambda_t} + b\left(\frac{\lambda_u}{\lambda_t}\right)^2
      \right] \; .
      \label{ratio2}
      \end{eqnarray}
 The real coefficients show a clear hierarchy $ c \gg a \gg b$ and the small ratio of CKM elements
 $\lambda_u/\lambda_t$ has an imaginary part.
 Thus in the SM the value of $\Delta \Gamma_s/\Delta M_s$ is well approximated by the term proportional to $c$,  while $a_{sl}^s$ is well approximated by the term proportional to $a$.
 We see from \eqref{ratio2} that the approximation $\lambda_u = 0 $ yields a vanishing semileptonic CP asymmetry. Since the current experimental uncertainty of $a_{sl}^s$ is about a factor of 130 larger than the SM central value of this quantity,  neglecting  $\lambda_u$  gives reasonable results.
 Moreover complex Wilson coefficients can have large effects in the semileptonic CP asymmetry by creating an imaginary part in the coefficient $c$, that does not suffer from the CKM suppression due to the factor  $\lambda_u/\lambda_t$.
   \subsection{The radiative decay \texorpdfstring{$B \to X_s \gamma$}{B to Xs gamma}}
 \label{sub:bsgamma}
\begin{figure}
\centering
\includegraphics[width=0.3\textwidth]{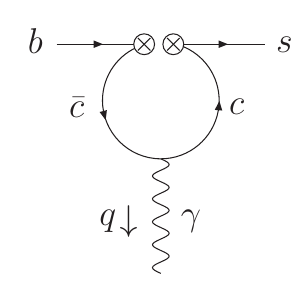}
\hspace*{1em}
\includegraphics[width=0.3\textwidth]{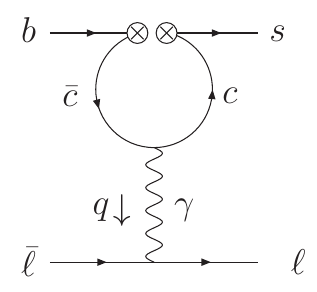}
\caption{Leading CSBM contribution to radiative decay (left) and rare \(b \to s \ell \ell\) decays (right).}
\label{fig:charm_loop_diagrams}
\end{figure}

 It is well known that weak radiative B meson decays are sensitive to BSM physics. The Standard-Model prediction of
 the branching ratio for  $\mathcal{B}(\bar{B}\rightarrow X_s\gamma)^\text{SM}=(3.36\pm0.23)\times 10^{-4}$ \cite{Misiak:2015xwa} is in
 good agreement with the current experimental average of $\mathcal{B}(\bar{B}\rightarrow X_s\gamma)^\text{exp}=(3.32\pm0.15)\times 10^{-4}$ \cite{Amhis:2016xyh}.
 In accordance with Heavy Quark Effective Theory (HQET) we may express the inclusive decay rate for a $B$ meson into a charmless hadron and a photon as
 \begin{equation}
 \Gamma(\bar{B}\rightarrow X_s\gamma)\simeq\Gamma(b\rightarrow X^\text{parton}_s\gamma)+\delta^\text{np},\\
 \end{equation}
 Here the non-perturbative term $\delta^\text{np}$, for $E_{\gamma}>E_0$ with the lower cut off of the photon energy
 $E_0=\SI{1.6}{\GeV}$, is estimated to be at the $(3\pm5)\%$ level \cite{Buchalla:1997ky,Benzke:2010js}.
Following the approach of \cite{Misiak:2006ab,Gambino:2001ew} the branching ratio
 $\mathcal{B}(\bar{B}\rightarrow X_s\gamma)$ can be expressed as
\begin{equation}
  \mathcal{B}(\bar{B}\rightarrow X_s\gamma)_{E_0>E_{\gamma}}=
  \mathcal{B}(\bar{B}\rightarrow X_c e\bar{\nu})^{\rm exp}
  \left\vert\frac{V_{ts}^*V^{}_{tb}}{V_{cb}}\right\vert^2
  \frac{6\alpha_{em}}{\pi C}
  \left[  P(E_0) + N(E_0)\right] \,,
  \label{eq:BrBXsgamma}
\end{equation}
where $P(E_0)$ and $N(E_0)$ denote, respectively, the leading-power perturbative contribution
and non-perturbative corrections, and C is defined as
\begin{equation}
C=\left\vert\frac{V_{ub}}{V_{cb}}\right\vert^2\frac{\Gamma(\bar{B}\rightarrow X_ce\bar{\nu})}{\Gamma(\bar{B}\rightarrow X_ue\bar{\nu})}\\.
\end{equation}
We neglect BSM corrections to the non-perturbative part and split the perturbative term $P(E_0)$ into
an SM part and a BSM part,
\begin{equation}
  P(E_0)  =   P^{\rm SM} (E_0) + P^{\rm BSM}(E_0)\label{eq:splitp0} .
  \end{equation}
We similarly split the branching ratio, 
\begin{equation}\label{eq:BR}
\mathcal{B}(\bar{B}\rightarrow X_s\gamma)=\mathcal{B}(\bar{B}\rightarrow X_s\gamma)_\text{SM}+\Delta\mathcal{B}_\text{BSM}\\.
\end{equation}
To zeroth order in $\alpha_s$ and neglecting the strange quark mass, we have
\begin{equation}\label{eq:deltap0}\
  P^\text{BSM}(E_0)  =  2C^{\, \rm eff, SM}_{7\gamma} \mathrm{Re}( \bar C_{7\gamma}^{\, \rm eff})
  +\vert \bar C_{7\gamma}^{\, \rm eff}\vert^2
  +\vert \bar C_{7\gamma}^{\, \prime \, \rm eff}\vert^2, \\
\end{equation}
which follows from substituting $C_{7\gamma}^{\rm eff} \to C_{7\gamma}^{\rm eff, SM} + \bar C_{7\gamma}^{\rm eff}$ in the SM expression.
The barred coefficients are defined as 
\begin{align}
\bar C_{7\gamma}^{\, \mathrm{eff}} (q^2,\mu)         &= C_{7\gamma}^{\mathrm{\, eff,BSM}}(\mu)
 + \frac{m_{c}}{m_{b}} \left[ 
\left( 4\Delta C_{9,10}^c (\mu) - \Delta C_{7,8}^c(\mu) \right) y(q^2,\mu) - \frac{\Delta C_{7,8}^c (\mu)}{6} \right] , \label{eq:C7bsm} \\
\bar C_{7\gamma}^{\, \prime \, \mathrm{eff}} (q^2,\mu)         &= C_{7\gamma}^{\, \prime \, \mathrm{eff,BSM}}(\mu)
+ \frac{m_{c}}{m_{b}} \left[ 
\left( 4\Delta C_{9,10}^{c\, \prime} (\mu) - \Delta C_{7,8}^{c\, \prime}(\mu) \right) y(q^2,\mu) - \frac{\Delta C_{7,8}^{c \, \prime} (\mu)}{6} \right] ,  
\label{eq:c7primebsm}&
\end{align}
where $\Delta C_{x,y}^c (\mu) = 3 \Delta C^c_x (\mu) + \Delta C^c_y (\mu)$ and
\begin{equation}
y(q^2, m_c, \mu) =  -\frac{1}{3} \left[ \ln \frac{m_c^2}{\mu^2} 
   - \frac32 + 2 a(z) \right] \,,
   \label{eq:yfunc}
\end{equation}
with $a(z)= \sqrt{|z-1|} \arctan \frac{1}{\sqrt{z -1}}$ and $z = 4 m_c^2/q^2$.  That is to say, in addition
to the BSM corrections to the Wilson coefficients $C_{7\gamma}^{(\prime) \, \mathrm{eff}}$, which arise from large logarithms
in the charm loop and are included through
leading-order RG evolution as described in Section \ref{sub:rge}, we also include the remainder of the
charm loop (see left diagram in Figure~\ref{fig:charm_loop_diagrams}).
For the SM contribution
and BSM contributions to $C_{1 \dots 4}^{c \, (\prime)}$, this is a constant of UV origin and is
already included in $C_7^{(\prime) \, \mathrm{eff, BSM}}$. In contradistinction, in the presence of nonzero CBSM
coefficients $C_{7 \dots 10}^{c \, (\prime)}$ a $q^2$-dependent contribution appears,
as is evident from the expressions. This comes together with a large logarithm at ${\cal O}(\alpha_s^0)$, which
causes a 1-loop mixing into $C_{7\gamma}^{\rm eff \, (\prime)}$, making the BSM contribution
formally ${\cal O}(1/\alpha_s)$ and leading over the SM,
as already mentioned in Section \ref{sub:rge}. In consequence, the finite contribution is formally subleading, as well as
scheme-dependent.\footnote{This scheme dependence will cancel against a corresponding scheme-dependence
in the (uncomputed) NLO correction to $C_7^{\rm eff \, (\prime)}$, which involves two-loop mixing
as well as the UV contribution to $C_{7\gamma}^{(\prime)}$}
However, the $q^2$ dependence itself
is a leading effect, and in any case is a qualitatively new feature, which is why we present it here.
For the coefficients $C_{5,6}^{c \, (\prime)}$, only a constant term is present and the mixing arises at two-loop
order, mirroring the situation of the SM.

With expressions \eqref{eq:splitp0}, \eqref{eq:BR} and \eqref{eq:deltap0} we determine the shift to the SM branching ratio in terms of the charmed four fermion coefficients and in Section
\ref{sec:pheno} find this leads, in many cases, to stringent constraints on their BSM parts $\Delta C_i^{c(\prime)}$.
For the numerics in this work we use $m_c = m_{c,\text{pole}}$ and work in the $q^2\rightarrow 0$ limit.
\subsection{Rare \texorpdfstring{$b \to s \ell \ell$}{b to s l l} decays}
\label{sub:bsll}
The rare decays $B \to K^{(*)} \ell^+ \ell^-$ are also calculable to leading order in the heavy-quark expansion \cite{Beneke:2001at}.
Through zeroth order in $\alpha_s$, they receive CBSM contributions both through
$\bar C_{7\gamma}^{(\prime) \, \mathrm{eff}}(q^2, \mu)$ defined in the previous subsection
and through further coefficients $C_9^{(\prime)\mathrm{eff}} (q^2, \mu)$ (see Figure~\ref{fig:charm_loop_diagrams} (right) for the contributing Feynman diagram). The CBSM contributions to the latter have the form
\begin{align}
 C_{9}^{\mathrm{eff, BSM}}(q^2, \mu) &= 
  C_{9V}^{\mathrm{BSM}}(\mu) +
\left(\Delta C_{1,2}^c (\mu) - \frac{\Delta C_{3,4}^c (\mu)}{2} \right) h(q^2, \mu) - \frac29 \Delta C_{3,4}^c (\mu) \, , \label{eq:C9}\\
 C_{9}^{\prime \mathrm{eff, BSM}}(q^2, \mu) &= 
  C_{9V}^{\prime \mathrm{BSM}}(\mu) +
\left(\Delta C_{1,2}^{c\prime} (\mu) - \frac{\Delta C_{3,4}^{c\prime} (\mu)}{2} \right) h(q^2, \mu) - \frac29 \Delta C_{3,4}^{c\prime} (\mu) \, , \label{eq:C9prime}
\end{align}
where
\begin{align}
 h(q^2, m_c, \mu) &= -\frac{4}{9} \left[ \ln \frac{m_c^2}{\mu^2} -
     \frac{2}{3}  + (2 + z) a(z) - z  \right] .
\end{align}
This $q^2$-dependent contribution to the rare $b \to s \ell \ell$ decays was a novel
feature in our previous work \cite{Jager:2017gal}, where we showed that, if the new physics scale is low enough,
it can be potentially observable. This contrasts with statements elsewhere in literature where such a $q^2$ dependence
was claimed to be an unambiguous criterion that the anomalies in the $B \to K^* \ell^+ \ell^-$ angular distribution
are of a hadronic origin. In the present work, we stick to new physics scales above the weak scale
(as in the ``high-scale scenario'' of \cite{Jager:2017gal}). In this situation, $C_{9V}$ is formally ${\cal O}(1/\alpha_s)$
already in the SM, although about half the numerical value originates from the
 ${\cal O}(1)$, formally subleading, $Wt$ loop.
Our $q^2$-dependent contributions are formally at the same order but turn out to be a numerically small correction.
The $q^2$-independent contribution to $C_{9V}^{(\prime)}$, however, can be dramatic.
In particular, when $\Delta C_1^{c}$ or $\Delta C_3^{c}$ are nonvanishing, it is easy to generate an ${\cal O}(1)$
shift to $C_{9V}$, while satisfying other constraints, as already stressed in \cite{Jager:2017gal}.
Similarly, $\Delta C_{1, 3}^{c \, \prime}$ strongly mix into $C_{9V}'$, which will allow us to set stringent constraints on
these coefficients in Section \ref{sec:pheno} below.

The scheme dependence of these effective coefficients enters through the $h$ and $y$ functions -- in general the constant
terms will vary depending on the choice of scheme.
In principle there is also a model-dependent contribution arising from the matching of the UV theory, which in this paper
we choose to ignore to keep our results model-independent.

When we consider the phenomenology in Section~\ref{sec:pheno}, we directly use $C^{\prime\,\mathrm{eff, BSM}}_{9}(q^2,\mu)$ and
$\bar{C}^{\prime\, \mathrm{eff}}_{7\gamma}(q^2,\mu)$ as ``observables''  and show the constraints they impose upon
the coefficients $\Delta C^{c\prime}_i$. Again, for the numerics in this work we use $m_c = m_{c,\text{pole}}$. We will take $q^2 = 0$ for $\bar C_{7\gamma}^{(\prime)}$,
 but will consider $C_9^{\prime\, \mathrm{eff, BSM}}$ at \(q^2 = \SI{5}{\GeV^2}\). This is because the former is
primarily constrained from polarisation observables in radiative decay and the low-$q^2$ end of the $B \to K^* \ell^+ \ell^-$ dilepton mass
distribution, while $C_9^{\prime\, \mathrm{eff,BSM}}$, if present, will become important away from the endpoint. We also note that
$C_9^{\prime\, \mathrm{eff}}$ and $C_7^{\prime \, \mathrm{eff}}$ are negligible in the SM.

 \subsection{The hadronic decay \texorpdfstring{$B_d \to J/ \psi K_S$}{Bd to J/psi KS}}
 \label{sub:BtoJpsiK}

 The $b \to c \bar{c} s$ operators listed in Section \ref{sec:setup} trigger the neutral $B$ meson decay
 $B_d\rightarrow J/\psi K_{S}$. As a peculiar feature, both $B_d$ and $\bar B_d$ can decay into the $J/\psi K_S$ final state, giving rise to a time-dependent CP-asymmetry via mixing-decay
interference. The amplitudes of these colour-suppressed tree-level decays read
\begin{eqnarray}
  \bar{\cal A}_{J/ \psi K_S} & = & \langle J/ \psi K_S | {\cal H}_{\rm eff} | \bar {B}_d \rangle
  = \frac{4\, G_F}{\sqrt{2}} \lambda_c \sum \limits_i C_i^c \langle J/\psi K_S | Q_i | \bar B_d \rangle \, ,
\\
 {\cal A}_{J/ \psi K_S} & = & \langle J/ \psi K_S | {\cal H}_{\rm eff} | B_d \rangle
  = - \frac{4 \,G_F}{\sqrt{2}} \lambda_c^* \sum \limits_i C_i^{c}{}^* \langle J/\psi K_S | Q_i | \bar B_d \rangle \, ,
\end{eqnarray}
where the hadronic matrix elements $\langle J/\psi K_S | Q_i | \bar B_d \rangle$ contain the (CP-even) strong rescattering
phases and the minus sign arises from $\eta_{\rm CP}(J/\psi K_S) = -1$. 
Their determination
is a non-perturbative problem which at present cannot be solved from first principles.
We will develop a strategy to extract partial information on these matrix elements, jointly with information on the
Wilson coefficients, from experiment in the following.

Let us express the branching ratio and time-dependent CP-asymmetry in terms of the Wilson coefficients and hadronic
matrix elements. Defining 
\begin{equation}
    \langle Q_i^c \rangle =  \langle J/ \psi K_S |Q_i^c | \bar B_d \rangle,
\qquad
    r_{i1}  =  \frac{\langle Q_i^c \rangle}{\langle Q_1^c\rangle} \in \mathbb{C} \, ,
\end{equation}
the branching ratio, obtained from the calculated $B^0_d\rightarrow J/\psi K^0_d$ decay rate along with  using the measured $B^0_d$ lifetime (for a similar approach see for example \cite{Cheng:1996xy}) reads
  \begin{eqnarray}
    \mathcal{B} (B_d \to J / \psi K_S) & = &
    \frac{\tau_{B_d} p_c G_F^2 \left| \lambda_c \right|^2}{M^2_{B_d} \pi}
    \left| \langle Q_1^c \rangle \right|^2
    \left| C^c_1 + C^c_2  r_{21} + C^c_3 r_{31} + C^c_4 r_{41} \right|^2 \, ,
    \label{branching}
    \nonumber
  \end{eqnarray}
where $p_c = 1.683$ GeV \cite{Tanabashi:2018oca} is the momentum of the final-state particles in the rest frame of the $B_d$
meson.

In writing \eqref{branching} we have omitted penguin contributions, which in what follows will be
negligible compared to the uncertainties stemming from the hadronic matrix elements, due to their small Wilson coefficients.

For the time-dependent CP-asymmetry, we neglect the tiny decay rate  difference $\Delta \Gamma_d$, such that it
takes the form
\begin{eqnarray}\label{eq:acp}
   A_{CP} (t)  &=&  \frac{\Gamma \left[ \bar{B}_d (t) \to\! J/\psi K_S \right] -  \Gamma \left[ {B}_d (t) \to\! J/\psi K_S \right]}
                           {\Gamma \left[ \bar{B}_d (t) \to\! J/\psi K_S \right] +  \Gamma \left[ {B}_d (t) \to\! J/\psi K_S \right]}
                           \nonumber
                           \\[3mm]
                           &=&  S_{J/\psi K_S} \sin(\Delta M_d t) - C_{J/\psi K_S} \cos(\Delta M_d t) \,  ,         
\end{eqnarray}
where $\Delta M_d$ is the precisely measured mass difference in the $B_d$ system.
Defining also
\begin{equation}
\label{eq:lambda}
 \lambda_{J/\psi K_S}  =  \frac{q_B}{p_B}  \frac{\bar{\cal A}_{J/\psi K_S}}{{\cal A}_{J/\psi K_S}} 
= - \frac{V_{tb}^* V_{td} }{V_{tb} V_{td}^* } \frac{\lambda_c}{\lambda_c^*}
           \frac{C_1^{c} + r_{21} C_2^{c} + r_{31} C_3^{c} + r_{41} C_4^{c}}
                 {C^{c*}_1 + r_{21} C^{c*}_2 + r_{31} C^{c*}_3 + r_{41} C^{c*}_4} \,  ,
\end{equation}
the mixing-induced asymmetry $S_{J/\psi K_S}$ and the direct CP asymmetry $C_{J/\psi K_S}$ are expressed as
\begin{equation}\label{eq:SandC}
 S_{J/\psi K_S} =\frac{2\,\mathrm{Im}\,\lambda_{J/\psi K_S}}{1+\vert\lambda_{J/\psi K_S}\vert^2} ,
\qquad      \qquad
 C_{J/\psi K_S}=\frac{1-\vert\lambda_{J/\psi K_S}\vert^2}{1+\vert\lambda_{J/\psi K_S}\vert^2}.
\end{equation}
In writing (\ref{eq:lambda}), we have neglected CP violation in the Kaon system and,
for the final expression, taken $q_B/p_B = \frac{V_{tb}^* V_{td}}{V_{tb} V_{td}^*}
= e^{-2 i \beta + {\cal O}(\lambda^4)}$,
which amounts to omitting CP-violation in mixing ($\lambda \approx 0.22$ is the Wolfenstein parameter). For reconstruction of \eqref{eq:acp}, \eqref{eq:lambda} and \eqref{eq:SandC} see for example \cite{Artuso:2015swg}, noting that in their notation $S_{J\psi K_S}=-\mathcal{A}_{CP}^{\mathrm{mix}}$ and $C_{J\psi K_S}=\mathcal{A}_{CP}^{\mathrm{dir}}$.

If the Wilson coefficients are real (recall their phase convention is fixed by (\ref{eq:Heff_bscc})),
the unknown hadronic matrix elements cancel out in (\ref{eq:lambda}) and, taking into account the tiny measured
value of $C_{J/\psi K_S}$, one obtains $S_{J/\psi K_S} \approx {\rm Im}(\lambda_{J/\psi K_S}) \approx \sin(2 \beta)$
to percent-level accuracy, as in the SM. Once the Wilson coefficients become complex, this is no longer true and
the $S$- and $C$-parameters become theoretically very uncertain. However, together with the branching fraction,
they still comprise three observables, allowing to jointly determine up to three parameters. Consider now a scenario
where new physics only affects $C_1^c$ or $C_2^c$. In this case, the only nonperturbative input to
the three observables is $| \langle Q_1^c \rangle |$ and $r_{21}$, comprising three real parameters in total. Hence
it is sufficient to have theoretical control over \textit{one} of these parameters in order to
obtain a constraint in the complex $C_1^c$ (or $C_2^c$) plane, generally a band.

 To proceed, let us use a Fierz-transformed basis
  \begin{eqnarray}
    O_1^c = (\bar{s}_L^i \gamma_\mu b_L^i) (\bar{c}_L^j \gamma^\mu c_L^j) \, ,
    &&
    O_2^c = (\bar{s}_L^i \gamma_\mu b_L^j) (\bar{c}_L^j \gamma^\mu c_L^i) \, ,
    \\
    O_3^c = (\bar{s}_L^i \gamma_\mu b_L^i) (\bar{c}_R^j \gamma^\mu c_R^j) \, ,
    &&
    O_4^c = (\bar{s}_L^i \gamma_\mu b_L^j) (\bar{c}_R^j \gamma^\mu c_R^i) \, ,
  \end{eqnarray}
  which relate to the operators defined in \eqref{eq:bscc_operator_basis} via
 \begin{eqnarray}
  Q_1^c=O_1^c \, , \, \, \, \,
  Q_2^c=O_2^c \, , \, \, \, \,
  Q_3^c=- \frac12 O_3^c \, , \, \, \, \,
  Q_4^c=- \frac12 O_4^c \, .
 \end{eqnarray}
The hadronic matrix element $\langle O_1^c \rangle = \langle Q_1^c \rangle$ factorizes naively in the limit of a large number
of colours. More precisely,
\begin{eqnarray}
   \langle O_1^c  \rangle & = &
       \langle J/\psi | \bar c_L^i \gamma^\mu | 0 \rangle \langle K_S | \bar s_L^i \gamma_\mu b_L^i | \bar B_d \rangle
 \left( 1 + {\cal O}(1/N_c^2) \right)
\\
&=& \frac{M_{B_d}}{2} 
   p_{c}  f_{J/ \psi} F^{B \to K} (q^2 = M_{J/ \psi}^2)  \left(1 + {\cal O}(1/N_c^2) \right),
\end{eqnarray}
resulting in an uncertainty of only ${\cal O}(10 - 20 \%)$ from form factor uncertainty and (nonfactorizable)
corrections to the $N_c \to \infty$ limit. Taking $F^{B \to K}(M^2_{J/\psi}) = 0.68 \pm 0.06$ (see Section \ref{sec:phenoBtopsiK}), one obtains
\begin{eqnarray}
 | \langle O_1^c \rangle | = \left[1.23 \pm 0.11 + {\cal O} \left( \frac{1}{N_c^2} \right) \right]  \mbox{GeV}^3
 = ( 1.23 \pm 0.18 )\, \mbox{GeV}^3 ,
 \end{eqnarray}
where in the last expression we have taken $1/N_c^2 = 1/9$ and combined errors in quadrature. 
This provides the required theoretical constraint. The three observables can now be used to jointly constrain the
complex Wilson coefficient and the complex matrix element ratio $r_{21}$. In a global fit, this amounts to determining
$r_{21}$ from data. This is the strategy which will be followed in our phenomenology below.

The fact that $r_{21}$ can (in the limited scenarios described) be determined from data motivates us to review
expectations for the other operator matrix elements. Let us write them as the sum of a naive-factorization result
and a correction term,
\begin{eqnarray}
    \langle J/ \psi K_S |O_i^c | \bar{B}_d \rangle & = & \frac{M_{B_d}}{2} 
   p_{c}  f_{J/ \psi} F^{B \to K} (q^2 = M_{J/ \psi}^2)
             \left[ r_0 + \sum \limits_{n=1}^{\infty} r_n \alpha_s^n + {\cal O} \left(  \frac{\Lambda_{\rm QCD}}{\alpha_s m_c} \right) \right] \, ,
             \nonumber \\
             \label{naive_factor}
 \end{eqnarray}
The constant $r_0$ amounts to the naive-factorization term, $r_0 = 1$ for $O_1^c$ and $O_3^c$ and  $r_0 =  1/3$ for $O_2^c$ and $O_4^c$.
QCD factorization in the heavy quark limit \cite{Beneke:2000ry} implies that the coefficients $r_n$, $n \geq 1$, are
perturbatively calculable but that there is also a power correction ``suppressed'' by only $\Lambda_{\rm QCD}/(\alpha_s m_c)$.
(See also \cite{Chay:2000xn,Cheng:2000kt}.)
As this power correction remains, at present, incalculable,
one does not expect predictivity, beyond the large-$N_c$ argument already
presented. From a large-$N_c$ perspective, the corrections to naive factorization are $1/N_c^2$-suppressed for both
$O_1$ and $O_3$, but are unsuppressed for $O_2$, $O_4$.

 The importance of contributions that do not naively
factorize (often called `non-factorizable') for these two operators becomes clearest by rewriting
 \begin{eqnarray}   \label{eq:singletoctet}
   O_2^c = \frac{1}{N_c} O_1^c + 2 T_1^c \, ,  &&  O_4^c = \frac{1}{N_c} O_3^c + 2 T_3^c \, ,
 \end{eqnarray}
 where
  \begin{eqnarray}
    T_1^c = (\bar{s}_L^i T^a _{ij}\gamma_\mu b_L^j) (\bar{c}_L^k T^a_{kl} \gamma^\mu c_L^l) \, ,
    &&
    T_3^c = (\bar{s}_L^i T^a_{ij} \gamma_\mu b_L^j) (\bar{c}_R^k T^a_{kl} \gamma^\mu c_R^l) \, .
  \end{eqnarray}
The matrix elements of the colour-octet operators $T_{1,3}^c$ vanish altogether in naive factorization, i.e. $r_0 = 0$ for them,
and they appear with large coefficients in (\ref{eq:singletoctet}).

Therefore the naive-factorization values
  \begin{eqnarray}
    r_{21}^{\rm (NF)} = \frac{1}{N_c} \, , \, \, \, \,
    r_{41}^{\rm (NF)} = \frac{1}{N_c} \, .
  \end{eqnarray}
should be expected to receive large corrections, whereas $r_{31} = 1 + {\cal O}(1/N_C^2)$. The situation is
further aggravated in the decay amplitude, which is proportional to the combination $C_1^c(\mu) + r_{21} C_2^c(\mu)$,
the so-called colour-suppressed tree amplitude which entails a severe cancellation in the SM case.

Quantitative estimates of $r_{21}$ beyond naive factorization have been obtained using QCD factorisation and/or light-cone
sum rules,   \cite{Melic:2003bw,Chay:2000xn,Cheng:2000kt}. In these approaches, the amplitude is usually parameterized as
$$
   {\cal N} \langle O_1^c \rangle^{\rm NF} a_2, 
$$
where ${\cal N}$ is a normalization and the scheme- and scale-independent combination $a_2$ is related to $r_{12}$ as
\begin{equation}
   a_2 = C_1^c(\mu) + r_{21}(\mu) C_2^c(\mu) + {\cal O}(\alpha_s^2) .
\end{equation}
For instance, the NLO QCD factorization results for $a_2$ in \cite{Cheng:2000kt} give
 $r_{21}(m_b) = 0.41 - 0.04 i$ or  $r_{21}(m_b) = 0.28 - 0.051 i$ for two different models of the $J/\psi$ light-cone distribution amplitudes. Both
numbers include an estimate of twist-three ($\Lambda/m_b$) power corrections but neglect the ${\cal O}(\Lambda/(\alpha_s m_c))$ corrections,
which as we have said provide an (additional) uncertainty.  By comparison, the LCSR computation in \cite{Melic:2003bw} finds values closer to
NF. These numbers can be contrasted to the `experimental' value $r_{21}(m_b) \sim 0.46$. It is important, however,
to note that this assumes the absence of NP in the Wilson coefficients $C_1^c$, $C_2^c$ (and in addition vanishing relative phase
between the hadronic matrix elements.) As we will see later, this assumption is not implied by
current experimental data when NP in $b \to c \bar c s$ transitions is allowed.

 \section{Phenomenology}
 \label{sec:pheno}
 \subsection{Numerical Inputs}
\label{sub:inputs}
In this section, we describe all the numerical inputs that are used in our work, along with the experimental results and their corresponding uncertainties.
We break these down into a set of fundamental inputs that are common to all our different observables, and then some specific input parameters for the
individual observables. Recall that quark masses are in the $\overline{\rm MS}$ scheme unless indicated otherwise.

\subsubsection{Common inputs}
We show in Table~\ref{tab:basic_inputs} input parameters that are common to all our theoretical predictions.
These inputs are taken from the Particle Data Group (PDG) \cite{Tanabashi:2018oca} and the CKMfitter
group \cite{Charles:2004jd}
(similar results for the CKM elements are also available from the UTfit collaboration \cite{Bona:2006ah}). 
\begin{table}
\centering
\begin{tabular}{l|l|l}
\hline
Parameter & Value & Reference \\
\hline
$\alpha_s (M_Z)$ & $0.1181(11)$ & PDG 2018 \cite{Tanabashi:2018oca}\\
$M_{K_S}$     & $0.497611(13)$ GeV & PDG 2018 \cite{Tanabashi:2018oca}\\
$M_{J/ \psi}$ &  3.096900(6) GeV & PDG 2018 \cite{Tanabashi:2018oca}\\
$M_{B_d}$     & 5.27955(26) GeV & PDG 2018 \cite{Tanabashi:2018oca}\\
$M_{B_s}$     & 5.36684(30) GeV & PDG 2018 \cite{Tanabashi:2018oca}\\
$\bar m_b \equiv m_b(m_b)$ & $4.18^{+0.04}_{-0.03}\,\si{\GeV}$ & PDG 2018 \cite{Tanabashi:2018oca}\\
$m_{c,\text{pole}}$ & 1.67(7) GeV & PDG 2018 \cite{Tanabashi:2018oca}\\
$m_c (\bar m_b)$ & 0.92(3) GeV & from $m_{c,\text{pole}}$ and $\alpha_s(M_Z)$ via RunDec \cite{Chetyrkin:2000yt,Herren:2017osy} \\
$m_s (\bar m_b)$ & $80^{+8}_{-6}\,\si{\MeV}$ & from $m_s(2 \,\rm GeV)$ \cite{Tanabashi:2018oca} and $\alpha_s(M_Z)$  via RunDec  \\
$|V_{ub} / V_{cb}|$ & $0.08835^{+0.00221}_{-0.00281}$ & CKMfitter \cite{Charles:2004jd} (ICHEP 2018 update) \\
$V_{cb}$ & $0.04240^{+0.00030}_{-0.00115}$ & CKMfitter \cite{Charles:2004jd} (ICHEP 2018 update) \\
$V_{us}$ & $0.2254745^{+0.000254}_{-0.000059}$ & CKMfitter \cite{Charles:2004jd} (ICHEP 2018 update) \\
$\gamma$ & $65.81^{+0.99}_{-1.66}\,\si{\degree}$ & CKMfitter \cite{Charles:2004jd} (ICHEP 2018 update) \\
\hline
\end{tabular}
\caption{List of input parameters needed for our theoretical predictions.}
\label{tab:basic_inputs}
\end{table}
 \subsubsection{Lifetime ratio}
The lifetimes of the $B_d$ and the $B_s$ meson are measured quite precisely. The Heavy Flavor Averaging
Group (HFLAV) quotes  \cite{Amhis:2016xyh}:
 \begin{eqnarray}
   \frac{\tau(B_s)}{\tau (B_d)}  =  0.993 \pm 0.004 \, ,
   &&
   \tau_{B_s}  =  \SI{1.509 \pm 0.004}{\ps}  \, .
    \end{eqnarray}
 The non-perturbative matrix elements of the $\Delta B=0$ operators were determined recently in \cite{Kirk:2017juj} with Heavy Quark Effective Field Theory
(HQET) sum rules,  using the following notation:
\begin{eqnarray}
  \langle B_{s}| Q_i | B_{s}\rangle = A_if^{2}_{B_{s}}M_{B_s}^2B_{i}\, ,
  &&
  \langle B_{s}| T_i | B_{s}\rangle =A_i f^{2}_{B_{s}}M_{B_s}^2\epsilon_{i} \, ,
\end{eqnarray}
with the coefficients
\begin{eqnarray}
  A_1=1 \, ,
  &&
  A_2=\frac{M_{B_s}^2}{(m_b+m_s)^2} \, .
\end{eqnarray}
We use the following numerical values for the decay constant from the Flavour Lattice Averaging Group
(FLAG)  \cite{Aoki:2016frl} and bag parameters from \cite{Kirk:2017juj}:
 \begin{eqnarray}
    f_{B_s}  = (227.2\pm3.4)\,   \mbox{MeV}, &&
    \\
   B_1(\bar{m}_b) = 1.028^{+0.064}_{-0.056}\,  ,
     &&
     B_2(\bar{m}_b)=0.988^{+0.087}_{-0.079}\, ,
      \\
      \epsilon_1(\bar{m}_b)=-0.107^{+0.028}_{-0.029}\, ,
      &&
      \epsilon_2(\bar{m}_b)=-0.033^{+0.021}_{-0.021}\, ,
      \end{eqnarray}
where $\bar m_b = m_b(m_b)$.
The tilded bag factors defined in Section~\ref{sub:lifetime} are expressed in terms of these as follows:
\begin{equation}
\widetilde{B}_{1,2} = 2 \epsilon_{1,2} + \frac{B_{1,2}}{N_c}
\end{equation}
For the SM value of the lifetime ratio we use the prediction from \cite{Kirk:2017juj}
\begin{eqnarray}
  \frac{\tau (B_s)}{\tau (B_d)} & = & 0.9994 \pm 0.0025 \, .
\end{eqnarray}
 \subsubsection{$B_s$ mixing}
 HFLAV quotes for the decay rate difference of $B_s$-mesons, $\Delta \Gamma_s$ and the semileptonic CP asymmetry $a_{sl}^s$  \cite{Amhis:2016xyh}
 \begin{eqnarray}
   \Delta \Gamma_s^{\rm exp}  =  \SI{0.088 \pm 0.006}{\ps^{-1}} \, ,
   &&
   a_{sl}^{s, \rm exp}         =  (- 60 \pm 280) \times 10^{-5}  \, .
      \end{eqnarray}
 For the $\Delta B=2$ matrix elements we use
 \cite{Aoki:2016frl,Artuso:2015swg}
 \begin{eqnarray}
f_{B_s} \sqrt{\hat{B}} = \SI{270(16)}{\MeV} \,,
\hspace{0.5cm}
\hat{B}= \num{1.32(6)} \, ,
\hspace{0.5cm}
\frac{\tilde{B}_S(\overline{m}_b)}{B(\overline{m}_b)}= \num{1.07(6)} \, .
 \end{eqnarray}
 The renormalization-group-invariant bag parameter $\hat{B}$ can be expressed in terms of the bag parameter $B$ from \eqref{MixME} as \cite{Aoki:2016frl}
\begin{equation}
\hat{B}=\left(\alpha_s(\mu)\right)^{-\frac{\gamma_0}{(2\beta_0)}}\left\{1+\frac{\alpha_s(\mu)}{4\pi}\left[\frac{\beta_1\gamma_0-\beta_0\gamma_1}{2\beta_0^2}\right]\right\}B(\mu)\\
\end{equation}
for $m_b < \mu < m_t $, and with $N_f=5$, and $N_c=3$, $\gamma_0=4,\beta_0=\frac{23}{3},\gamma_1=\frac{116}{3},\beta_1=-\frac{43}{9}$.
We follow  \cite{Artuso:2015swg} and use 
\begin{equation}
\hat{B}=1.51599 \, B(m_b) \, . 
\end{equation}
For the SM value of the mixing observables we use \cite{Artuso:2015swg}
\begin{eqnarray}
  \Delta \Gamma_s^{\rm SM}  =  \SI{0.088 \pm 0.02}{\ps^{-1}} \, , &&   a_{sl}^{s, \rm SM}  =  \num{2.22 \pm 0.27 e-5} \, .
\end{eqnarray}
\subsubsection{\texorpdfstring{$B \to X_s \gamma$}{B to Xs gamma}}
For the inclusive radiative $B_s$ meson decay we use the experimental average obtained by HFLAV \cite{Amhis:2016xyh} and the
 theoretical prediction from Misiak et al. \cite{Misiak:2015xwa}: 
 \begin{eqnarray}
 \mathcal{B}(\bar{B}\rightarrow X_s\gamma)^\text{exp}&=&(3.32\pm0.15)\times10^{-4}\,,
 \\
  \mathcal{B}(\bar{B}\rightarrow X_s\gamma)^\text{SM}&=&(3.36\pm0.23)\times10^{-4}.
 \end{eqnarray}
  The semileptonic branching ratio and the ratio $C$ as defined in Section~\ref{sub:bsgamma} are taken from \cite{Misiak:2006ab}:
 \begin{eqnarray}
   \mathcal{B}(\bar{B}\rightarrow X_c e\bar{\nu})^\text{exp} & = &  0.1061 \pm 0.0017 \, ,
   \\
   C   & = & 0.580 \pm 0.016 \, .
 \end{eqnarray}
The SM contribution to $B \to X_s \gamma$ which interferes with our BSM contribution is given by \cite{Czakon:2015exa}
\begin{equation}
C_{7\gamma}^\text{eff,SM} (m_b) = -0.385  \, .
\end{equation}
    
 \subsubsection{Rare decays from BSM operators}

 In light of the recent anomalous measurements of $b \to s \ell \ell$ decays by LHCb, in particular the $R_{K^{(*)}}$ results
 \cite{Aaij:2014ora,Aaij:2017vbb,Aaij:2019wad} there has been considerable work on fitting the semileptonic and radiative Wilson  coefficients to data.
 While $R_{K^{(*)}}$ are indicative of a lepton-flavour-non-universal NP effect, a UV 
completion of the EFT will, in general, also include a lepton-flavour-universal effect. Such a combined
scenario has been shown to be consistent with (and even mildly preferred by) the data in \cite{Geng:2017svp}
(see also \cite{Alguero:2018nvb,Alguero:2019ptt,Aebischer:2019mlg}).
In \cite{Jager:2017gal}, we have studied the possible $C_{9V}$ effects
generated by $C_1^c \dots C_4^c$ in detail. 
In the presence of the operators $C_i^{c\prime}$ involving right-handed strange quarks, the CBSM
scenario produces an effect in $C'_{9V}$ as well, associated with the
 right-handed semileptonic operator $Q'_{9V}$. We will treat $C^{\prime\mathrm{eff,BSM}}_{9}(q^2,\mu)$ in \eqref{eq:C9prime} as a pseudo-observable
and use the below value taken from the fit in \cite{Aebischer:2019mlg} to constrain our model
\begin{equation}
 C_9^{\prime\, \mathrm{eff, BSM, exp}} = \num{0.09 \pm 0.15} \, ,
\end{equation}
where the theoretical uncertainties associated with exclusive rare $B$ decay are included in this value by the authors \cite{Aebischer:2019mlg} and are taken from \cite{Altmannshofer:2014rta,Straub:2015ica,Straub:2018kue} compatible with \cite{Khodjamirian:2010vf,Bobeth:2017vxj}.
Similarly, rare and radiative $b$ decays can receive CBSM contributions via the coefficient
$C_{7\gamma}'$ of the right-handed dipole operator $Q'_{7\gamma}$. We treat $\bar{C}_{7\gamma}^{\prime\,\mathrm{eff}}(q^2,\mu)$ in \eqref{eq:c7primebsm} as a second
pseudo-observable, with our experimental value taken from the fit in
\cite{Paul:2016urs},
\begin{equation}
\bar C^{\prime\,\mathrm{eff, exp}}_{7\gamma} = \num{0.018 \pm 0.037} .
\end{equation}
(Some more recent fits \cite{Alguero:2019ptt,Arbey:2019duh} give very similar results for the \(\bar C^{\prime\,\mathrm{eff, exp}}_{7\gamma}\) coefficient that would not change our numerics significantly.)
Recall that we take $q^2 = 5 \mbox{GeV}^2$ in $C_9^{\prime\,\mathrm{eff, BSM}}$ and $q^2 = 0$ in $\bar C_{7\gamma}^{\prime\,\mathrm{eff}}$.
We have identified $C_9^{\prime\,\mathrm{eff}} = C_9^{\prime\,\mathrm{eff, BSM}}$ and
$C_{7\gamma}^{\prime\,\mathrm{eff}} = \bar C_{7\gamma}^{\prime\,\mathrm{eff}}$due to the negligible SM contributions.

  \subsubsection{Observables in \texorpdfstring{$B_d \to J / \psi K_S$}{Bd to J/psi KS}}
\label{sec:phenoBtopsiK}
For the decay $B_d \to J/\psi K_S$ we take the CP violating observables
\begin{eqnarray}
\label{eq:hflav_S_C}
  S_{J/\psi K_S} = \num{0.699 \pm 0.017}
  \,,
  &&
  C_{J/\psi K_S} = \num{-0.005 \pm 0.015} \,,
\end{eqnarray}
from \cite{Amhis:2016xyh}, and the branching ratio
\begin{equation}
\mathcal{B} (B_d\rightarrow J/\psi K_S) = \num{8.73 \pm 0.32 e-4} \,,
\end{equation}
from \cite{Tanabashi:2018oca}.
As part of our theoretical calculation of $S_{J/\psi K_S}$ and $C_{J/\psi K_S}$, we use the most recent CKMfitter \cite{Charles:2004jd} value
\begin{equation}
\label{eq:CKMfitter_sin2beta}
\sin 2\beta = 0.738^{+0.027}_{-0.030} \,,
\end{equation}
where the experimental measurement is not included in their fit.
We note there is a very slight tension between the HFLAV average and the CKMfitter result, at the level of \(\sim 1.1\,\sigma\).

For the non-perturbative decay constant of the $J / \psi$ resonance,
we take the value from the phenomenological study in \cite{Bailas:2018car}
\begin{eqnarray}
  f_{J/ \psi} & = & (407 \pm 6) \, \mbox{MeV} \, .
  \end{eqnarray}
This value agrees well with the lattice determinations in \cite{Becirevic:2013bsa,Donald:2012ga,Bailas:2018car}.
The form factor can be determined via LCSR \cite{Khodjamirian:2017fxg} or extrapolated from lattice simulations
\cite{Bailey:2015dka,Bouchard:2013pna}. Both approaches have similar uncertainties, the LCSR values are slightly larger than
the lattice results.
\begin{align}
  F^{B \to K} (q^2 = M_{J/ \psi}^2) & = \num{0.68 \pm 0.06} \quad \text{LCSR} \, ,
  \\
  F^{B \to K} (q^2 = M_{J/ \psi}^2) & = 0.59 \pm 0.06 \quad \text{lattice (Fermilab/MILC)} \,,
  \\
  F^{B \to K} (q^2 = M_{J/ \psi}^2) & = 0.59 \pm 0.06 \quad \text{lattice (HPQCD)} \,.
  \end{align}
Since the LCSR value is a direct evaluation at \(q^2 = M_{J/\psi}^2\), we have chosen that as our input for the evaluation of the \(\langle O_1^c\rangle\) matrix element.
However we note for the reader that using an average instead would give very similar results, albeit with a slightly smaller central value and error, and therefore our choice is conservative.

\subsection{Constraints on $\Delta C_5-\Delta C_{10}$ and $\Delta C_1^{\prime}-\Delta C_{10}^{\prime}$}
\label{sec:realcoefs}
In this section, we consider constraints upon real Wilson coefficients of operators $Q^c_5-Q^c_{10}$ and primed operators  $Q^{\prime c}_1-Q^{\prime c}_{10}$. To determine confidence intervals for individual Wilson coefficients, we switch on
BSM contributions in one Wilson coefficient at a time and set all others to their SM values.
We construct a \(\chi^2\) test statistic from the experimental measurements of our chosen observables and our theoretical predictions, combining the experimental and theoretical errors in quadrature\footnote{We make the assumption that the experimental and theoretical errors are Gaussian distributed and independent.}:
\begin{equation}\label{eq:chi2individ}
\chi^2_i(\vec{C}^c)=\frac{\left(\mathcal{O}_i^\text{th}(\vec{C}^c)-\mathcal{O}^\text{exp}_i\right)^2}{(\sigma^\text{exp}_i)^2+(\sigma^\text{th}_i)^2} \,,
\end{equation}
where the index \(i\) runs over $\mathcal{B}(\bar{B}\rightarrow X_s\gamma)$, $\Delta \Gamma_s$ and $\frac{\tau (B_s)}{\tau (B_d)}$ and in addition our ``pseudo-observables'' $\bar{C}_{7\gamma}^{\prime\mathrm{eff}}(q^2,\mu)$ and $C^{\prime \mathrm{eff,BSM}}_{9}(q^2,\mu)$.
The $1\sigma$ intervals implied by individual observables are displayed in Tables \ref{tab:individ} and \ref{tab:angular}.
To obtain combined constraints (Table \ref{tab:lambdas}), we sum up the individual $\chi^2$.
In all cases we normalise to the best fit value by subtracting the relevant $\chi^2$ value at the minimum,
$\Delta \chi^2 = \chi^2 - \chi^2_{\rm min}$.
\begin{table}[ht]
\begin{tabular}{|m{1cm}|m{4.8cm}|m{4.9cm}|m{4.3cm}|}
\hline
Coeff.& $\Delta\chi^2_{\gamma}\leq1$&$\Delta\chi^2_{\tau}\leq1$& $\Delta\chi^2_{\Delta\Gamma}\leq1$\\
\hline
  $\Delta C_ 5$ & $[-0.01,0.01],[0.36,0.37]$ & $[-0.03,-0.01],[0.03,0.06]$ & $[-0.13,0.34]$ \\
  \hline
 $\Delta C_ 6$ & $[-0.02,0.03],[1.1,1.2]$ & $[-0.11,-0.03],[0.09,0.17]$ & $[-1.5,0.49]$ \\
 \hline
 $\Delta C_ 7 $& $[-0.46,-0.45],[-0.01,0.01]$ & $[-0.21,-0.11],[0.04,0.14]$ & $[-1.7,0.44]$ \\
 \hline
 $\Delta C_ 8 $& $[-0.92,-0.88],[-0.02,0.014]$ & $[-0.26,-0.12],[0.06,0.20]$ & $[-0.27,0.27]$ \\
 \hline
 $\Delta C_ 9$ & $[-0.002,0.003],[0.15,0.15]$ & $[-0.02,-0.01],[0.003,0.011]$ & $[-0.14,0.035]$ \\
 \hline
 $\Delta C_{10}$ & $[-0.05,0.07],[3.2,3.3]$ & $[-0.08,-0.05],[0.02,0.05]$ & $[-0.09,0.09]$ \\
 \hline
 $\Delta C^{\prime}_{1}$ & $[-5.7,5.7]$ & $[-0.32,-0.15],[0.08,0.25]$ & $[-0.58,0.58]$ \\
 \hline
  $\Delta C^{\prime}_{2}$ & $[-0.53,0.53]$ & $[-1.2,-0.51],[0.39,1.1]$ & $[-0.39,0.39]$ \\
 \hline
  $\Delta C^{\prime}_{3}$& $[-6.7,6.7]$ & $[-1.0,-0.79],[0.06,0.30]$ & $[-1.1,1.1]$ \\
 \hline
 $\Delta C^{\prime}_{4}$& $[-0.75,0.75]$ & $[-1.3,-0.96],[0.09,0.45]$ & $[-0.44,0.44]$ \\
 \hline
 $\Delta C^{\prime}_{5}$& $[-0.05,0.05]$ & $[-0.03,-0.01],[0.03,0.06]$ & $[-0.21,0.21]$ \\
 \hline
 $\Delta C^{\prime}_{6}$& $[-0.15,0.15]$ & $[-0.10,-0.03],[0.10,0.18]$ & $[-0.85,0.85]$ \\
 \hline
 $\Delta C^{\prime}_{7}$& $[-0.06,0.06]$ & $[-0.23,-0.13],[0.03,0.13]$ & $[-0.86,0.86]$ \\
 \hline
  $\Delta C^{\prime}_{8}$& $[-0.12,0.12]$ & $[-0.30,-0.17],[0.04,0.17]$ & $[-2.0,2.0]$ \\
 \hline
 $\Delta C^{\prime}_{9}$& $[-0.02,0.02]$ & $[-0.02,-0.01],[0.003,0.011]$ & $[-0.07,0.07]$ \\
 \hline
 $\Delta C^{\prime}_{10}$& $[-0.42,0.42]$ & $[-0.09,-0.05],[0.01,0.05]$ & $[-1.2,1.2]$ \\
 \hline
\end{tabular}
\caption{$1\sigma$ intervals for scenarios with one Wilson coefficient. }
\label{tab:individ}
\end{table}
\begin{table}[ht]
\begin{center}
\begin{tabular}{|m{1cm}|m{4.8cm}|m{4.9cm}|}
\hline
Coeff.& $\Delta\chi^2_{C_9^{\prime\mathrm{eff,BSM}} }\leq1$&$\Delta\chi^2_{\bar{C}_{7\gamma}^{\prime\mathrm{eff}}}\leq1$\\
\hline
 $\Delta C^{\prime}_{1}$ & $[-0.01,0.02]$ & $[-1.10,3.2]$ \\
 \hline
 $\Delta C^{\prime}_{2}$ & $[-0.02,0.11]$ & $[-0.30,0.10]$ \\
  \hline
 $\Delta C^{\prime}_{3}$ & $[-0.04,0.01]$ & $[-3.7,1.3]$ \\
 \hline
 $\Delta C^{\prime}_{4}$ & $[-0.08,0.02]$ & $[-0.42,0.14]$ \\
 \hline
 $\Delta C^{\prime}_{5} $& $-$& $[-0.01,0.03]$ \\
 \hline
 $\Delta C^{\prime}_{6} $& $-$& $[-0.03,0.08]$ \\
 \hline
 $\Delta C^{\prime}_{7} $& $-$ & $[-0.03,0.01]$ \\
 \hline
 $\Delta C^{\prime}_{8}$ & $-$ & $[-0.07,0.02]$ \\
 \hline
 $\Delta C^{\prime}_{9}$ & $-$ & $[-0.004,0.01]$ \\
 \hline
 $\Delta C^{\prime}_{10}$ & $-$ & $[-0.08,0.23]$ \\
 \hline
 \end{tabular}\caption{$1\sigma$ intervals for scenarios with one primed Wilson coefficient. These correspond to allowed ranges for $ \bar{C}_{7\gamma}^{\prime \mathrm{eff}}$ and $C_9^{\prime\mathrm{eff,BSM}} $ at $\pm 1\sigma$.}
\label{tab:angular}
\end{center}
\end{table}

Considering $\Delta C_5-\Delta C_{10}$ in the first column of Table \ref{tab:individ}, there are best fit ranges which correspond to those that pass through the SM point and those that do not. This can be understood by considering the functional form with which $\bar{C}_{7\gamma}^{\mathrm{eff}}$ enters \eqref{eq:deltap0} and the impact that larger contributions from coefficients in $C_{7\gamma}^\text{eff, BSM}$ have upon reducing the parameter space allowed by radiative decay. The second column of Table \ref{tab:individ} does not for any coefficient include the SM point, and this is simply due to the current disagreement between measurement and theory for the lifetime ratio. Column 3 containing ranges accommodated by $\Delta\Gamma_s$ always includes the SM point. In Table \ref{tab:angular} we show $1 \sigma$ ranges for the primed coefficients, accommodated by pseudo observables $C_{9}^{\prime\mathrm{eff,BSM}}$ and $\bar{C}_{7\gamma}^{\prime\mathrm{eff}}$ . 
In the second column of Table \ref{tab:lambdas} we show the $1\sigma$ allowed ranges for $\Delta C_5-\Delta C_{10}$ which follow from combining constraints of all observables. The $1\sigma$ ranges for the primed coefficients $\Delta C_1^{\prime}-\Delta C_{10}^{\prime}$ however exclude the radiative decay branching ratio observable, as this is already contained in the fitted value of $\bar C_{7\gamma}^{\prime\mathrm{eff,exp}}$.
In the last two columns of Table \ref{tab:lambdas} we re-express the combined bounds in terms of the scale of new physics $\Lambda_{\rm NP}$, defined through
\begin{equation}\label{eq:lambdanp2}
   \frac{4G_F}{\sqrt{2}} |V_{cb}V^*_{cs}| |\Delta C_{i}| =\frac{1}{\Lambda_{\rm NP}^2} .
\end{equation}
The lower bound on the NP scale corresponding to the negative boundary of
the $1\sigma$ interval for $\Delta C_i$ is denoted as $\Lambda_-$, that corresponding to the positive boundary as
$\Lambda_+$. For $\Delta C_{10}$ and $\Delta C_9'$ the $1\sigma$ interval only contains positive values and only
$\Lambda_+$ is given, corresponding to the upper boundary. For $\Delta C_{10}'$, the $1\sigma$ region is composed
of two intervals, and $\Lambda_-$ and $\Lambda_+$ correspond to the smallest and largest Wilson coefficient values
($-0.08$ and $0.05$, respectively).
\begin{table}[ht]
\begin{center}
\begin{tabular}{|m{1cm}|m{5.2cm}|m{1.5cm}|m{1.5cm}|}
\hline
Coeff.& $\Delta\chi^2 \leq 1$&$\Lambda_- (\si{\TeV})$& $\Lambda_+ (\si{TeV})$\\
\hline 
 $\Delta C_5$ & $[-0.01,0.01]$ & 9.7 &10.5 \\
 \hline 
 $\Delta C_6$ & $[-0.02,0.02]$ & 5.6 &5.8 \\
 \hline 
 $\Delta C_7$ & $[-0.01,0.01]$ & 8.8& 9.7 \\
 \hline 
 $\Delta C_8$ & $[-0.02,0.02]$ & 6.2 & 6.9 \\
 \hline 
 $\Delta C_9$ & $[-0.001,0.01]$ & 22.3 & 12.6  \\
 \hline 
 $\Delta C_{10}$ & $[0.01 , 0.05]$ & - & 3.8  \\
 \hline 
 $\Delta C^{\prime}_{1}$ & $[-0.01 ,0.02]$ &11.9&5.5 \\
 \hline 
 $\Delta C^{\prime}_{2}$ & $[-0.04, 0.09]$ & 4.5& 2.8 \\
 \hline 
 $\Delta C^{\prime}_{3}$ & $[-0.04, 0.02]$ &4.5&7.0 \\
 \hline 
 $\Delta C^{\prime}_{4}$ & $[-0.07 , 0.03]$ & 3.2 &5.1 \\
 \hline 
 $\Delta C^{\prime}_{5}$ & $[-0.02 ,0.04]$ & 5.8&4.5  \\
 \hline 
 $\Delta C^{\prime}_{6}$ & $[-0.07 , 0.11]$ & 3.3	&2.6  \\
 \hline 
 $\Delta C^{\prime}_{7}$ & $[-0.03 ,0.02]$ & 5.1 &6.6 \\
 \hline 
 $\Delta C^{\prime}_{8}$ & $[-0.06 , 0.04]$ & 3.6	&4.3\\
 \hline 
 $\Delta C^{\prime}_{9}$ & $[0.002,0.010]$ &-&8.5 \\
 \hline 
 $\Delta C^{\prime}_{10}$ & $[-0.08 , -0.06],[0.02 , 0.05]$ & 3.1	&3.8  \\
 \hline
\end{tabular}\caption{Allowed $1\sigma$ ranges from all observables combined and corresponding bounds on BSM scale.}
\label{tab:lambdas}
\end{center}
\end{table}
We see that with our definition of $\Lambda_{\rm NP}$, which is agnostic to the details of
BSM physics generating the $\Delta C_i^c$,
our observables can provide sensitivity to scales higher than 20 TeV in scenarios involving
the tensor Wilson coefficient $\Delta C^{(\prime)}_{9}$. Other Wilson coefficients probe somewhat lower scales, but
always at least 3 TeV.

For scenarios in which we consider NP in pairs of Wilson coefficients, setting all others to their SM values; the pattern of constraints may be divided into three categories:
\begin{enumerate}[label=(\roman*)]
\item Coefficients $\Delta C_5-\Delta C_{10}$:
As explained in Section \ref{sub:rge}, the mixing of operators $Q^c_5,...,Q^c_{10}$ with $Q_{7\gamma}$ first at one-loop gives rise to strong RG effects which enter $\bar{C}_{7\gamma}^{\mathrm{eff}}$ and result in the dominant constraint for such scenarios coming from radiative decay. As the coefficient $\bar{C}_{7\gamma}^{\mathrm{eff}}$ enters \eqref{eq:deltap0} both quadratically as well as linearly, it receives a contribution from $C_{7\gamma}^{\mathrm{eff, SM}}$. This combination results in a much narrower 1 sigma region, as is shown in all panels of Figure \ref{fig:reunprimed} as the blue shaded area. In the first and third panel, the presence of another purple band corresponds to contours where $\bar{C}_7^{\mathrm{eff}}=-2 C_{7\gamma}^{\mathrm{eff},SM}$. In terms of the lifetime ratio, shown in the green shaded area, the contours slightly miss the SM point due to the current $1.4\sigma$ discrepancy between theory and experiment. For the $B_s-\bar{B}_s$ width difference the scenarios consisting of coefficients of operators with left and right handed vector currents and coefficients of tensor operators are the most restrictive. In all cases scenarios between even numbered coefficients are favoured owing to the $\frac{1}{N_c}$ suppression which always accompanies colour singlet operators in the calculations.
\begin{figure}[ht]
\begin{subfigure}{0.3\textwidth}
\includegraphics[width=\textwidth]{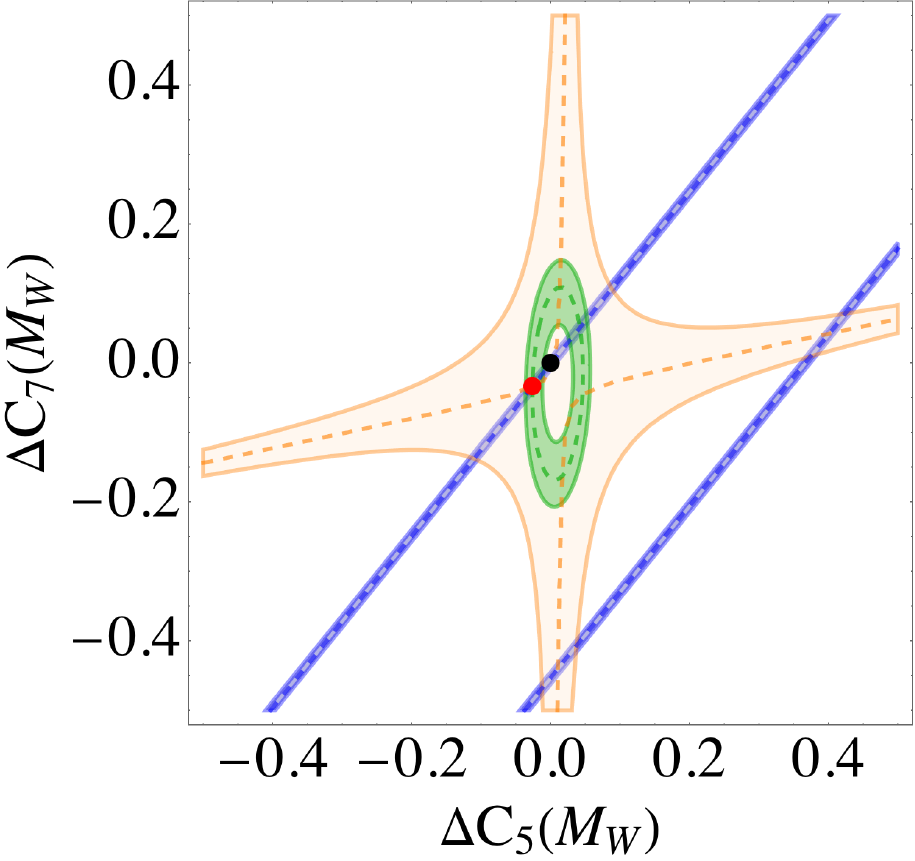}
\end{subfigure}
\begin{subfigure}{0.3\textwidth}
\includegraphics[width=\textwidth]{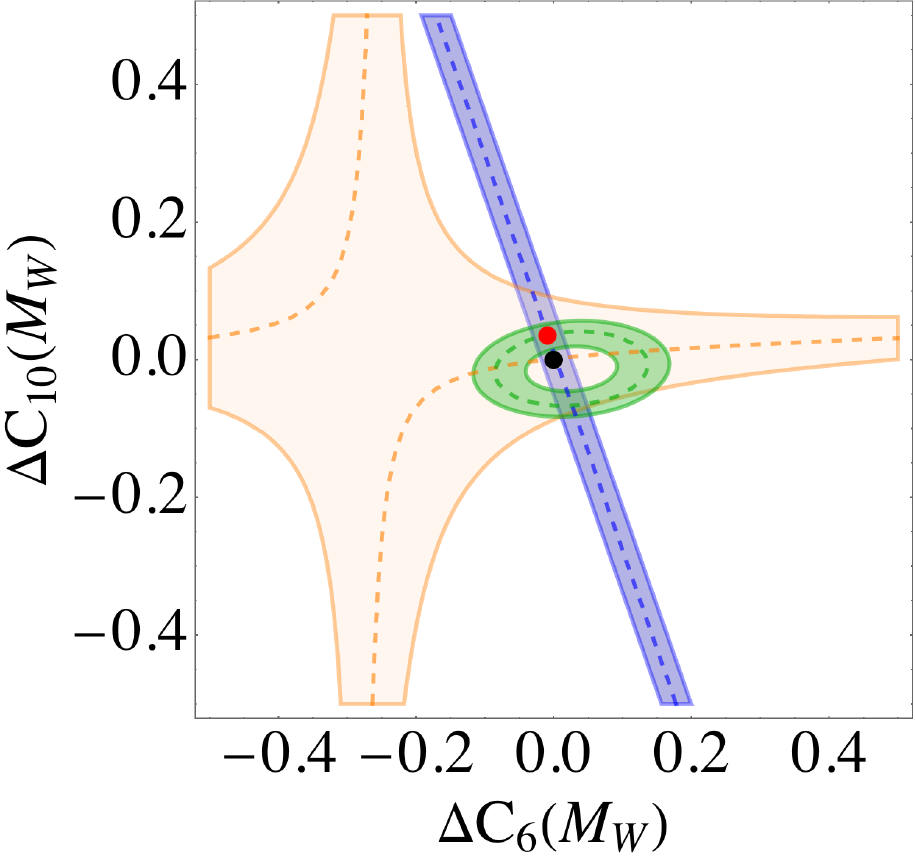}
\end{subfigure}
\begin{subfigure}{0.3\textwidth}
\includegraphics[width=\textwidth]{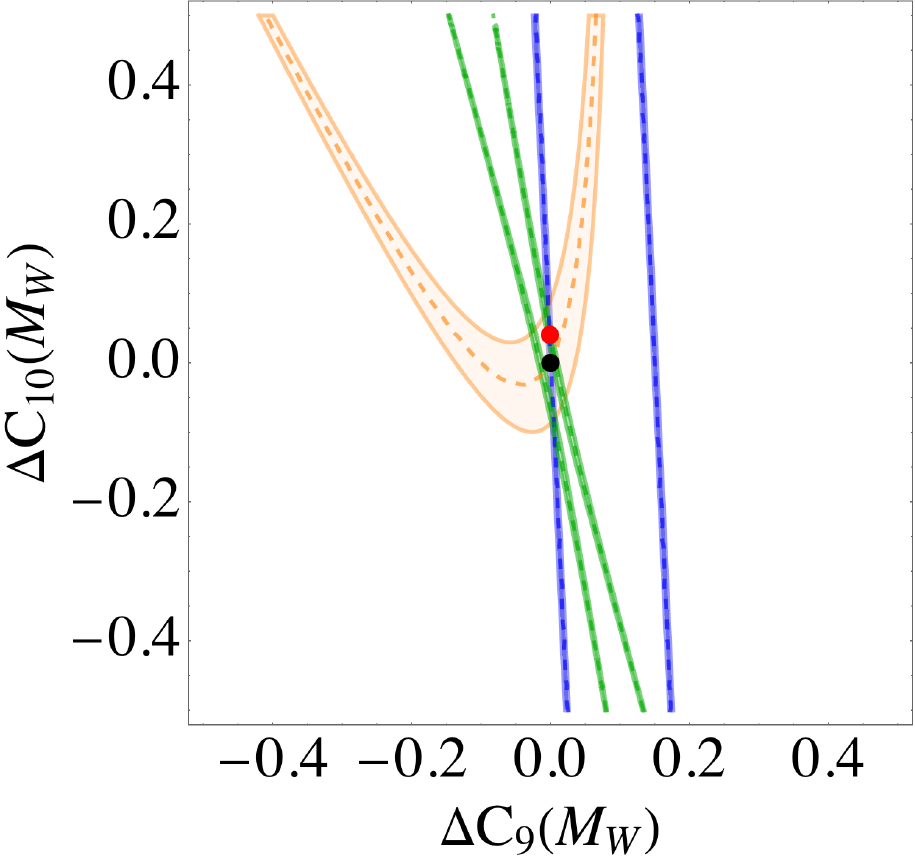}
\end{subfigure}
\caption{Overlaid individual constraints from radiative decay (blue), lifetime ratio (green), width difference (orange) upon $\Delta C_5-\Delta C_7$ plane (left), $\Delta C_6-\Delta C_{10}$ plane (middle), $\Delta C_9-\Delta C_{10}$ plane (right). The SM point and best fit point are shown as the black and red dots respectively.}\label{fig:reunprimed}
\end{figure}
\item $\Delta C^{\prime}_1-\Delta C^{\prime}_{4}$:
The 6 plots in Figure \ref{fig:primed} show as additional constraints, contours of $C_{9}^{\prime\mathrm{eff,BSM} }=0.09\pm0.15$ and of  $\bar{C}_{7\gamma}^{\prime \mathrm{eff}}=0.018\pm0.037$ in 2 parameter planes of Wilson coefficients $\Delta C^{\prime}_{1}(M_W)-\Delta C^{\prime}_{4}(M_W)$. Here the central value used is the best fit points for $C_{9V}^{\prime}$ and $C_{7\gamma}^{\prime}$ acquired from global fits to angular observables in \cite{Aebischer:2019mlg}, \cite{Paul:2016urs}. These are to be compared with the high scale scenarios considered already in \cite{Jager:2017gal} and discussed in Section \ref{sub:c1to4}. For combinations of these coefficients it is found that the strongest constraint comes from the experimental fits of angular observables to $C_{9V}^{\prime}$. This is due to a strong dependence of $C_{9V}^{\prime \text{BSM} }(m_b)$ upon $\Delta C^{\prime}_{1}(M_W)-\Delta C^{\prime}_{4}(M_W)$
which results in closely spaced contours (red dashes). The inclusive radiative decay branching ratio constraint (blue shading) is shown for extra information, in addition to that given by the contours of $\bar{C}^{\prime\mathrm{eff}}_{7\gamma}$ at fitted value of $\bar C_{7\gamma}^{\prime\mathrm{eff,exp}}$ (which includes inclusive branching ratio data). For these primed coefficients the radiative decay constraint is very much weaker than in the previous case due to the mixing of operators $Q_1^{c\prime},...,Q_4^{c\prime}$ with $Q^{\prime}_{7\gamma}$ occurring first at two-loop, and in addition, to the primed coefficient $\bar{C}_{7\gamma}^{\prime \mathrm{eff}}$ only entering \eqref{eq:deltap0} quadratically with no linear dependence upon $C_{7\gamma}^{\mathrm{eff, SM}}$.
\begin{figure}
\begin{subfigure}{0.3\textwidth}
\includegraphics[width=\textwidth]{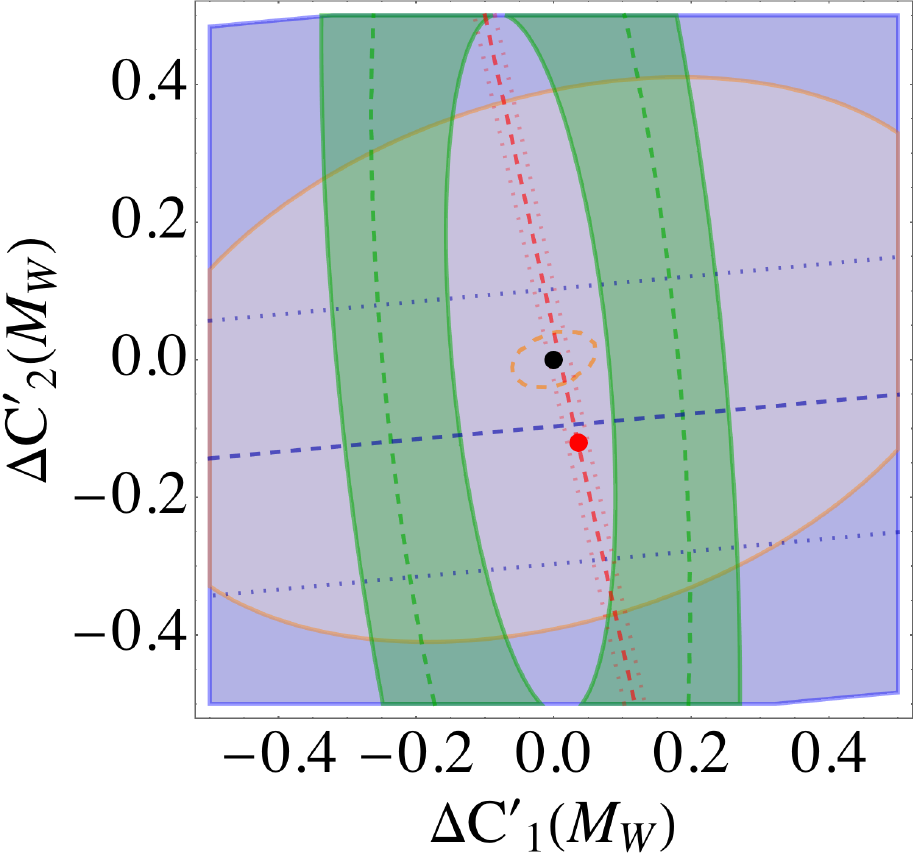}
\end{subfigure}
\hspace*{\fill}
\begin{subfigure}{0.3\textwidth}
\includegraphics[width=\textwidth]{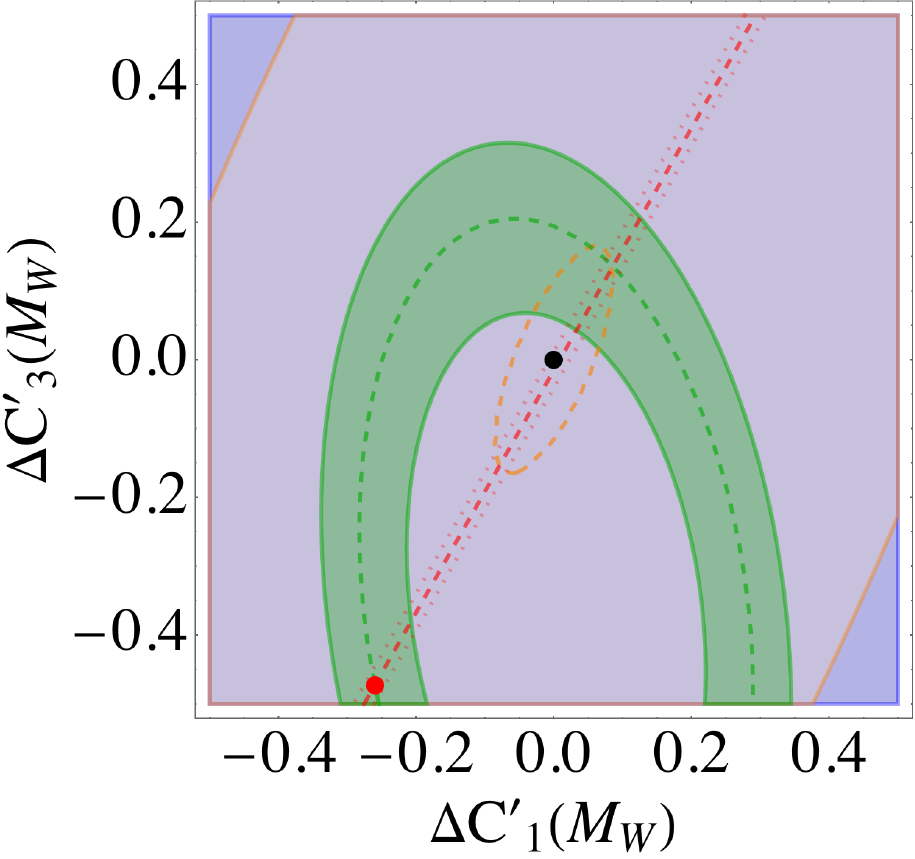}
\end{subfigure}
\hspace*{\fill}
\begin{subfigure}{0.3\textwidth}
\includegraphics[width=\textwidth]{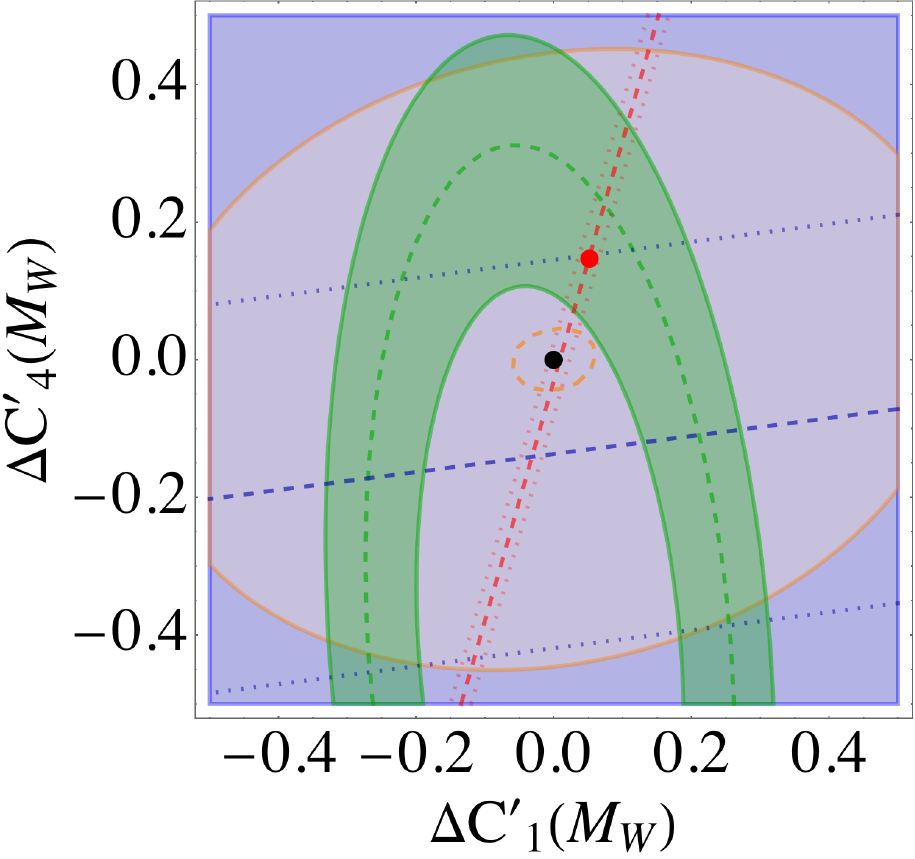}
\end{subfigure}
\begin{subfigure}{0.3\textwidth}
\includegraphics[width=\textwidth]{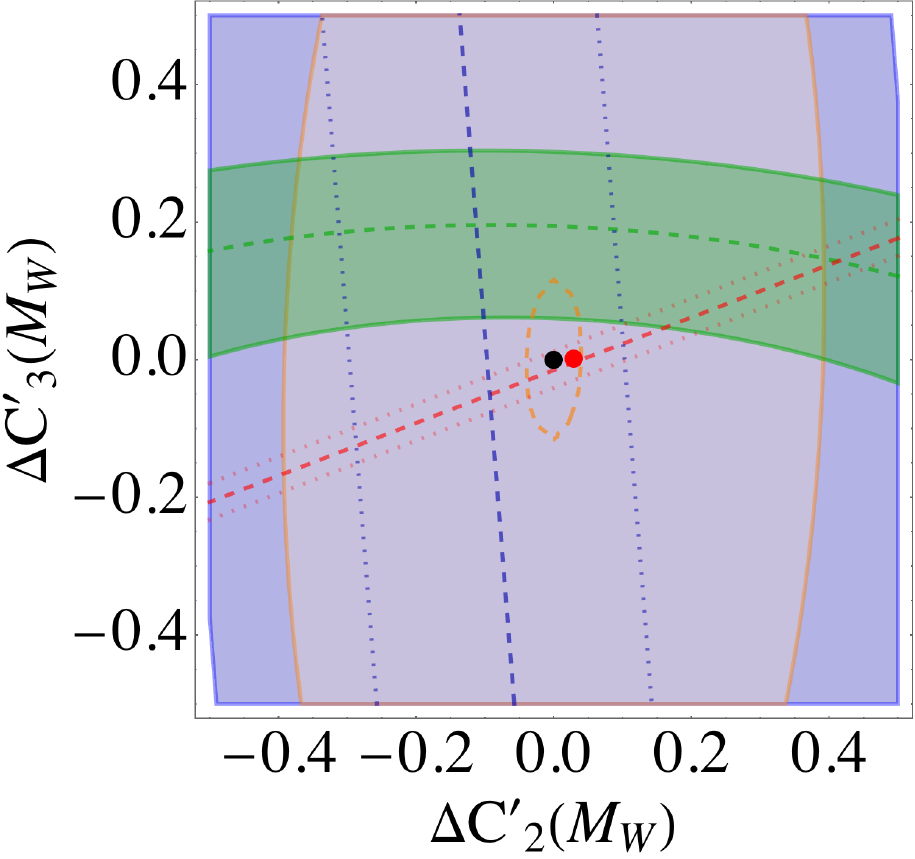}
\end{subfigure}
\hspace*{\fill}
\begin{subfigure}{0.3\textwidth}
\includegraphics[width=\textwidth]{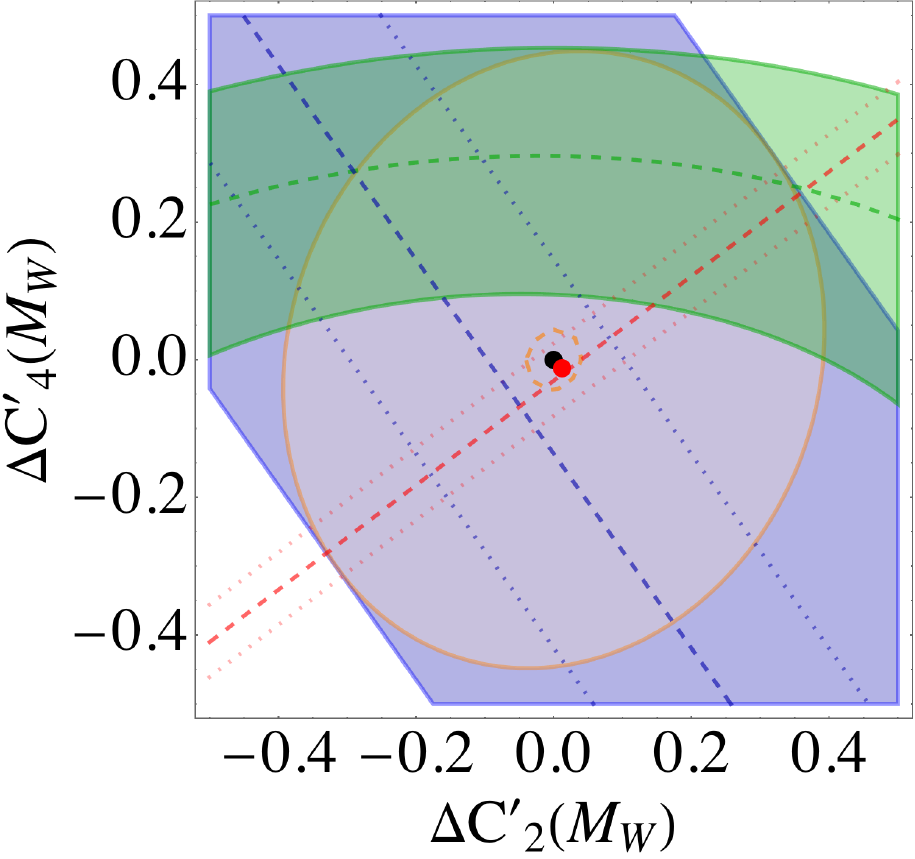}
\end{subfigure}
\hspace*{\fill}
\begin{subfigure}{0.3\textwidth}
\includegraphics[width=\textwidth]{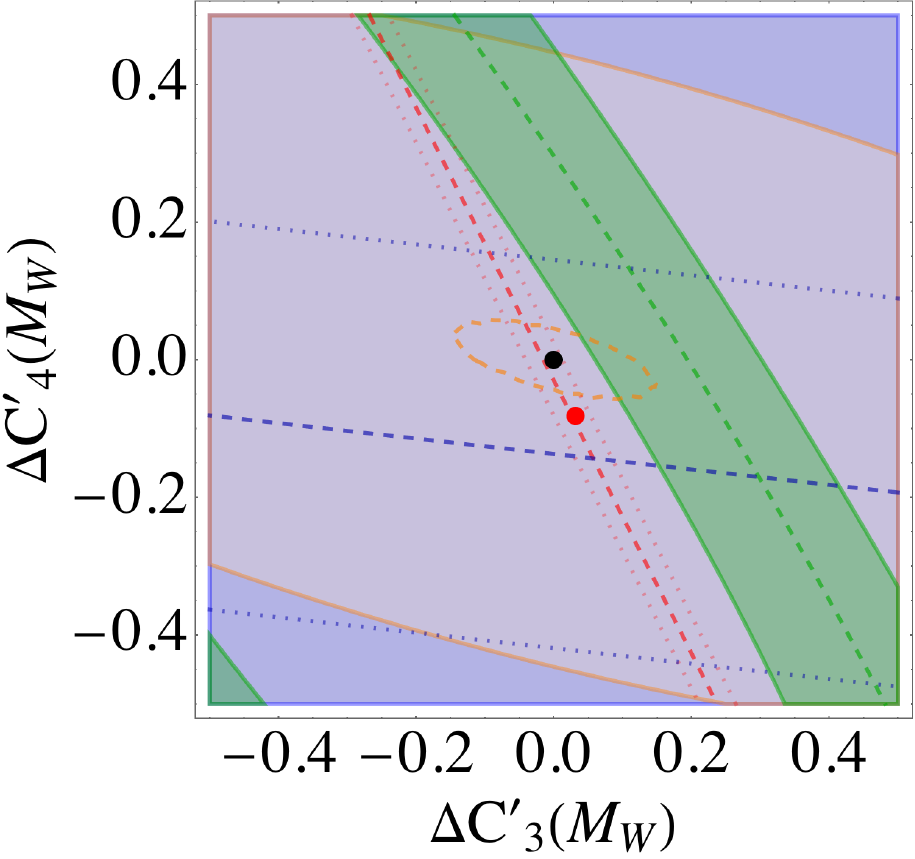}
\end{subfigure}
\caption{Contours of $C_{9}^{\prime\mathrm{eff,BSM} }=0.09\pm0.15$  (red, dashed) and of  $\bar{C}_{7\gamma}^{\prime \mathrm{eff}}=0.018\pm0.037$ (blue, dashed), along with radiative decay (blue), lifetime ratio (green) and width difference (orange) \(1 \sigma\) constraints upon two parameter scenarios involving Wilson coefficients $\Delta C^{\prime}_1-\Delta C^{\prime}_4$. The SM point and best fit point are shown as the black and red dots respectively.}\label{fig:primed}
\end{figure}
\item $\Delta C^{\prime}_5-\Delta C^{\prime}_{10}$:
Three examples of possible 2 parameter scenarios are shown in Figure \ref{fig:reprimed} and are the right handed counter parts in one to one correspondence with those of Figure \ref{fig:reunprimed}. As in (i) the one-loop mixing under renormalization of  $Q^{c\prime}_5,...,Q^{c\prime}_{10}$ with $Q^{\prime}_{7\gamma}$ results in a stronger dependence of $\bar{C}_{7\gamma}^{\prime \mathrm{eff}}$ upon $\Delta C^{\prime}_{5}-\Delta C^{\prime}_{10}$ and this results in strict constraints upon combinations of these coefficients. All plots show very similar constraints from the lifetime ratio and the width difference, although differences are more pronounced in $\Delta \Gamma_s$ due to primed and unprimed coefficients not mixing in the theoretical prediction for $\Gamma_{12}^{c\bar{c}}$ and hence there is no linear contribution from SM parts of Wilson coefficients $C^c_1$ and $C^c_2$ here. The best fit points shown in red are all placed close to the SM point (black), but are pulled away slightly by the lifetime ratio constraint.
\end{enumerate}
\begin{figure}
\begin{subfigure}{0.3\textwidth}
\includegraphics[width=\textwidth]{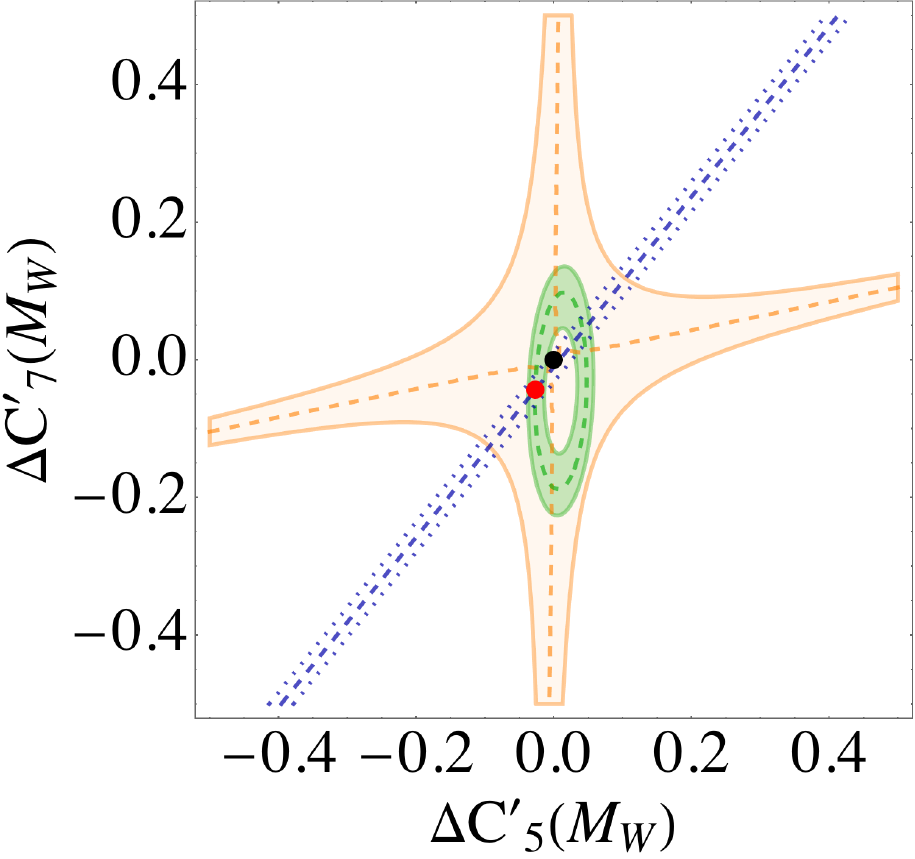}
\end{subfigure}
\begin{subfigure}{0.3\textwidth}
\includegraphics[width=\textwidth]{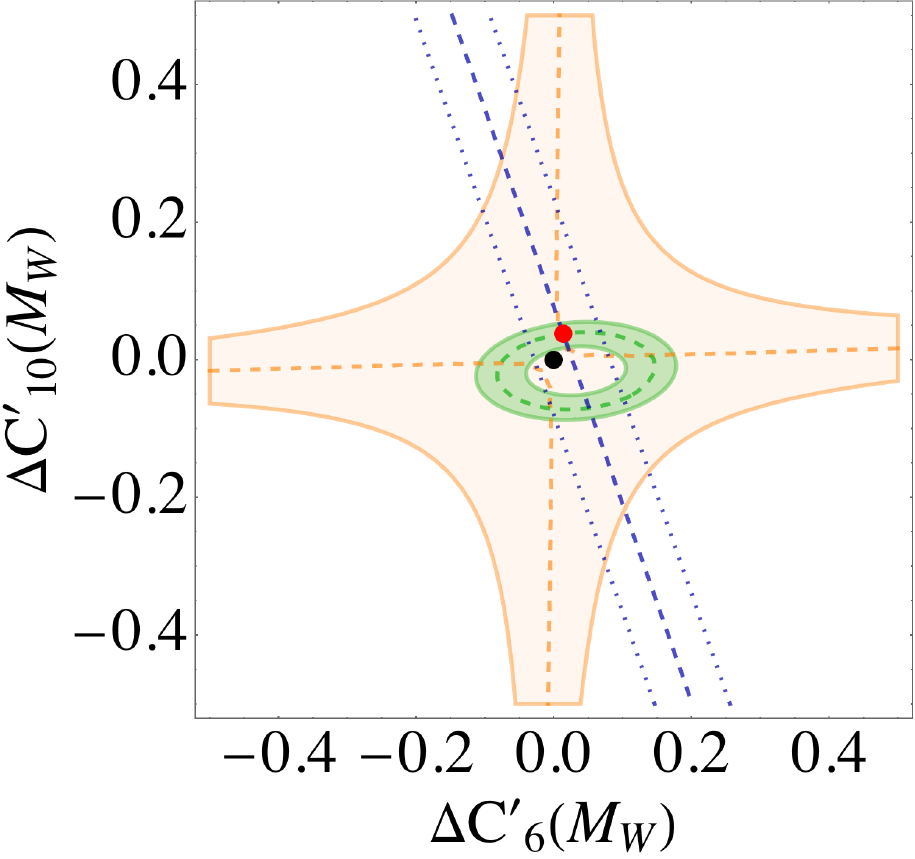}
\end{subfigure}
\begin{subfigure}{0.3\textwidth}
\includegraphics[width=\textwidth]{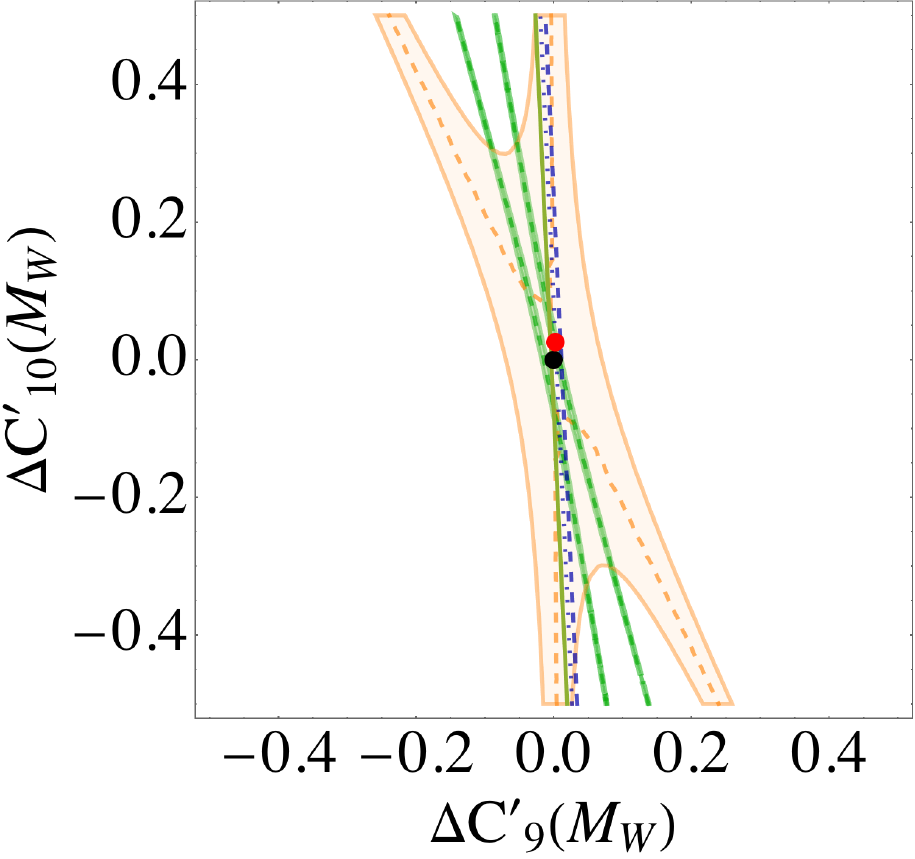}
\end{subfigure}
\caption{Overlaid individual constraints upon $\bar{C}_{7\gamma}^{\prime\mathrm{eff}}$ (blue dashed central value/dotted $\pm\sigma$), lifetime ratio (green), width difference (orange) upon $\Delta C^{\prime}_5-\Delta C^{\prime}_7$ plane (left), $\Delta C^{\prime}_6-\Delta C^{\prime}_{10}$ plane (middle), $\Delta C^{\prime}_9-\Delta C^{\prime}_{10}$ plane (right). The SM point and best fit point are shown as the black and red dots respectively.}\label{fig:reprimed}
\end{figure}

\subsection{The case of $\Delta C_1 - \Delta C_4$}
\label{sub:c1to4}

For the case of the \(C^c_{1-4}\), the phenomenology for a scenario with purely real coefficients was covered in our previous paper \cite{Jager:2017gal} -- we
briefly recap our conclusions from that work, before expanding to a scenario with complex coefficients and the constraints that arise from \(B_d \to J/\psi K_S\) decays.
Complex NP in \(C^c_1\) and \(C^c_2\) is studied in the recent work \cite{Lenz:2019lvd} where they compute constraints arising from \(\tau(B_s)/\tau(B_d)\) and \(\mathcal{B}(B \to X_s \gamma)\) as part of a sophisticated global fit. 
We performed a mutual check with the authors of that work, and found full agreement on the constraints arising from the previously mentioned observables (i.e. the blue and green bands in Figure~\ref{fig:nf_allowed_complex_c1_2} below).

In \cite{Jager:2017gal}, we studied NP confined to the first four operators of the full basis \eqref{eq:bscc_operator_basis}, as these operators give a large contribution to a (lepton flavour universal) shift in the \(C_{9V}\) coefficient (since the RG running coefficients are \(\mathcal{O}(1/\alpha_s)\) in the logarithmic counting), while only being constrained by the radiative decays through two-loop RG mixing.
In our study we found that while the SM is consistent with the lifetime ratio, width difference and radiative decay observables, there is also room for a shift in our \((\bar c b)(\bar s c)\) Wilson coefficients without disagreement with data -- see Fig.\ 3 in \cite{Jager:2017gal}. For instance,
a shift to the \(C^c_3\) coefficient alone of order 0.2 could produce a shift of $C_{9V}$ of order $-1$, and such an NP contribution is in fact slightly favoured with respect to the Standard Model, as it lessens the small tension present in \(\tau(B_s) / \tau(B_d)\).
NP in two coefficients simultaneously can also be accommodated, such as in
the pair \((\Delta C_2, \Delta C_4) = (-0.1, 0.3)\) which generates an \(\mathcal{O}(1)\) contribution to \(C_{9V}\) (albeit with no change relative to the SM in the lifetime ratio).
In light of the fact that current data supported a possible NP contribution in several different Wilson coefficients, the natural question was how to distinguish between these scenarios.
We showed that an improvement in the future precision of both the lifetime ratio and the width difference \(\Delta \Gamma_s\) could identify the particular realisation of charming BSM physics in nature.

As introduced in Section~\ref{sub:BtoJpsiK}, NP in \((\bar c b)(\bar s c)\) operators can alter the \(B_d \to J/\psi K_S\) decay rate, as well as the related CP asymmetries \(S_{J/\psi K_S}\) and \(C_{J/\psi K_S}\) in the case of complex Wilson coefficients.
In order to predict these three observables, hadronic operator matrix elements
$\langle J/\psi K_S | O_i^c | \bar B \rangle$
must be evaluated.
In the SM, the NF approximation for them does not give good agreement data, and it is widely
assumed that deviations from NF can bring the prediction in line with experiment;
as reviewed in Section~\ref{sub:BtoJpsiK}, large deviations from naive factorization are
theoretically to be expected for $\langle O_2^c \rangle$ but not $\langle O_1^c \rangle$.

However, in this section we jointly consider the three observables, to constrain either $\Delta C_1$ or
$\Delta C_2$ together with the uncertain matrix element ratio $r_{21}$ defined
in Section~\ref{sub:BtoJpsiK}, with the matrix element $\langle O_1 \rangle$, for which
violations of NF are expected to be small, taken in the range described there. Constructing a
$\chi^2$ test statistic out of the three observables and the $\langle O_1 \rangle$ range and
profiling it over $\langle O_1 \rangle$ and $r_{21}$ results in a constraint in the complex $\Delta C_{1,2}$
planes. 

For the complex \(\Delta C_1\) plane, the resulting constraint is shown in
 Figure~\ref{fig:nf_allowed_complex_c1_2} (left) as red bands, implying
a remarkably powerful constraint from this hadronic decay: the imaginary part must
be close to either zero or \(\pm 0.2\). In effect, we have determined the uncertain matrix element
ratio $r_{21}$ from data. We show in the figure the 1, 2, and \(3\,\sigma\) regions for the $B_d \to J/\psi K_S$ constraints.
The $\mathrm{Im }\,C_1 = 0$ band is very narrow, and has no $1 \sigma$ region because of the slight discrepancy for $\mathrm{sin}2\beta$ mentioned in Section~\ref{sec:phenoBtopsiK}.
In the same figure we overlay the constraints from $a_{sl}^s$ (yellow), $\Delta \Gamma_s$ (orange), and \(\tau(B_s) / \tau(B_d)\) (green). We see that the lifetime ratio is the most constraining 
among the three, while the other two rule out some regions with larger
real and/or imaginary NP contributions ($B \to X_s \gamma$ provides no visible constraints at this scale).
Combined with the constraint from $B \to J/\psi K_S$, only a few small regions in the $\Delta C_1$
plane are allowed.

In fact, the $\chi^2$-profiling allows, at each point in the plane, to simultaneously determine the
value of $r_{21}$ that gives the best fit. For reasons of computational simplicity, we carry this out only
along a circle in the middle of the (green) lifetime ratio ring, where the experimental central
value of \(\tau(B_s) / \tau(B_d)\) is obtained. Along this ring, the (complex) value of $r_{21}$ varies
substantially. Along the two short black segments it is in excellent agreement with
naive factorization. Surprisingly, these two segments happen to lie in two of the small regions in
the plane that are allowed by all constraints. We have no explanation for this curious fact.
But it certainly demonstrates that experimental data in $B \to J/\psi K_S$ does \textit{not} imply
large violations of naive factorization, in contrast to a widely held belief. (Similar results are
found when varying the radius of the circle within the lifetime band, corresponding to different
fixed values of \(\tau(B_s) / \tau(B_d)\).)

\begin{figure}
\centering
\includegraphics[width=\textwidth]{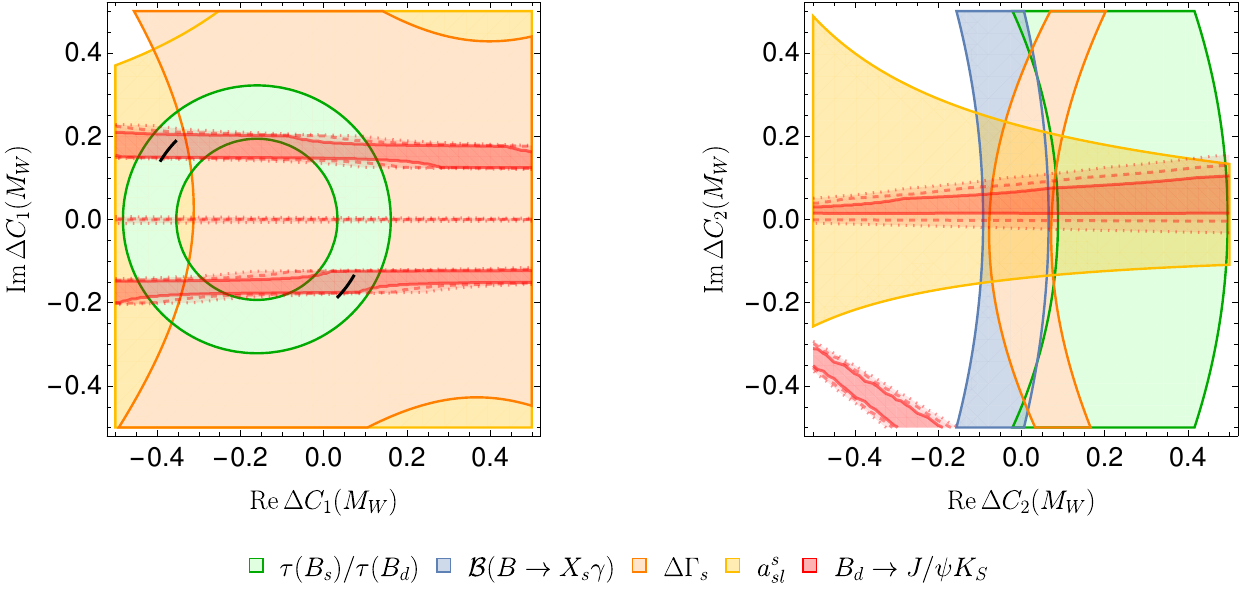}
\caption{
  Left: Constraints from $B_d \to J/\psi K_S$ and other constraints on the complex $\Delta C_1$ plane.
  The red horizontal bands uses a theory prediction only for $\langle O_1 \rangle$, expected to be
close to its naive-factorization value as reviewed in Section~\ref{sub:BtoJpsiK}. Each
band shows regions of agreement at the \(1\sigma\) (solid), \(2\sigma\) (dashed), and \(3 \sigma\) (dotted) levels. (Note that the band lying on the real axis only has 2 and \(3 \sigma\) agreement, due to the small tension between \eqref{eq:hflav_S_C} and \eqref{eq:CKMfitter_sin2beta}).
On the black arcs, NF for both \(\braket{O_1^c}\) and \(r_{21}\) is in agreement with data.
  Right: Same, but for NP in $\Delta C_2$.
}
\label{fig:nf_allowed_complex_c1_2}
\end{figure}

The results of repeating the study for complex \(\Delta C_2\) are displayed in Figure~\ref{fig:nf_allowed_complex_c1_2} (right). In this case,  \(B_d \to J/\psi K_S\) data allows a band centred on the real, as well as a second band with negative real and imaginary parts that however is ruled out by the
other constraints; among those, the most stringent constraints now come from the lifetime difference
$\Delta \Gamma_s$ and the radiative decay $B \to X_s \gamma$.
Extending beyond the SM operators $Q_{1,2}^c$ to NP in $Q_{3,4}^c$, the theoretical prediction for $B_d \to J/\psi K_S$ requires more non-perturbative inputs in the form of \(r_{31}, r_{41}\), alongside \(r_{21}\) (as there is still a SM contribution to \(C^c_2\)).
Attempting to fit these from the $B_d \to J/\psi K_S$ data does not bring any further insight, as there is sufficient freedom to always explain the measurements.

\section{Conclusions}
\label{sec:conclusions}

In this paper we have made a thorough study of the possible effects of new physics arising in tree-level \(b \to c \bar{c} s\) decays.
This partonic decay mode contributes to a wide variety of different observables. Among them, the branching ratio for radiative $B$ meson decay \(\mathcal{B} (B \to X_s \gamma)\), the $B$ meson lifetime ratio \(\tau(B_s) / \tau(B_d)\), and the $B_s$ mixing observables \(\Delta \Gamma_s\) and \(a_{sl}^s\) stand out:
They are inclusive decay modes that are well measured experimentally, and are theoretically controlled through the HQE.
In addition, we have shown that, in the CP-violating case,
new constraints which are only midlly affected by theoretical uncertainties arise from exclusive $B_d \to J/\psi K_S$ observables.
Taken together, effects in this set of observables are correlated in our ``Charming BSM'' scenario, and the observables
provide very complementary constraints on it.

The space of NP contributing to \(b \to c \bar{c} s\) decays is spanned by 20 operators, defined in \eqref{eq:bscc_operator_basis}.
We have calculated the contribution from the full basis to all of our observables; the most complex result being that obtained for mixing and the lifetime ratio for which we used tools from the Mathematica package FeynCalc \cite{Mertig:1990an,Shtabovenko:2016sxi}. Our full results are given in \eqref{eq:gamma12bsm}--\eqref{eq:gamma12G} and \eqref{eq:gamcalc}--\eqref{eq:gammaijrules} (these results are also available as ancillary \texttt{Mathematica} files with arXiv version of this article).
We have further calculated the renormalization group evolution for our basis. In the SM case we used known results for ADM entries available in the literature \cite{Bobeth:2003at,Chetyrkin:1996vx} and for the mixing of our operators as described in Section~\ref{sub:rge}, elements from \cite{Buras:2000if}, our previous work \cite{Jager:2017gal}, as well as those elements extracted from our $b\rightarrow s\ell\ell$ results either directly or through substitution of relevant colour factors. Our results are summarised in the full evolution matrix given in \eqref{eq:Utot}.
An explicit recipe for making use of our results to place constraints on an arbitrary NP model is as follows:
\begin{enumerate}
  \item Match the chosen BSM model onto our effective Hamiltonian (given in \eqref{eq:Heff_bscc}) at the scale \(M_W\).
  \item Use the RG evolution matrix (given in \eqref{eq:Utot}) to run the effective coefficients down to the scale \(m_b\).
  \item Use the low scale Wilson coefficients as inputs to the algebraic expressions for the observables \(\Delta \Gamma_s\) (given in \eqref{eq:gamma12bsm}--\eqref{eq:gamma12G}, or the attached file \texttt{Gamma12.m}), \(\tau(B_s) / \tau(B_d)\) (given in \eqref{eq:gamcalc}--\eqref{eq:gammaijrules}, or the attached file \texttt{LifetimeRatio.m}), and \(\mathcal{B} (B \to X_s \gamma)\) (given in \eqref{eq:BrBXsgamma}--\eqref{eq:yfunc}), to generate BSM theory predictions in terms of the BSM theory parameters.
  \item Compare the computed BSM theory predictions to the corresponding experimental measurements of choice. 
\end{enumerate}

We have extended our earlier results \cite{Jager:2017gal} in two important ways.
Firstly, we complement our previous analysis of real new physics in \(Q_{1-4}^c\) by studying the possibility of CP-violating NP, and introduced observables from  \(B_d \to J/\psi K_S\) as constraints. Secondly, we studied constraints from our main set of observables on the Wilson coefficients \(C_{5-10}^{c(\prime)}\) and in analogy with our study of \(C_{1-4}^{c}\) we have constrained from global fits to the ``wrong chirality'' coefficients \(C'_{9V}\) and \(C'_{7\gamma}\), possible BSM effects in  \(C_{1-4}^{c(\prime)}\).

When considering the introduction of new weak CP violating phases to SM coefficients, we have used the sine and cosine coefficients  \(S_{J/\psi K_S}\) and \(C_{J/\psi K_S}\) of the time dependent CP asymmetry, alongside the branching ratio \(\mathcal{B}(B_d \to J/\psi K_S)\) to constrain the parameter space. By treating the uncertain hadronic matrix element ratio \(r_{21} = \braket{O^c_2}/\braket{O^c_1} \) as a free parameter and profiling over it, we
effectively determine it from data, reducing the required theoretical input to the colour-allowed matrix element  \(\braket{O^c_1}\). We assume this to
 be close to its value in naive factorization, as is expected from large-$N_C$ counting. In this way we obtained constraints relying only mildly
on theory, even though we are dealing with nonleptonic exclusive decays.

For NP in \(\Delta C_1\), our result (shown in Figure~\ref{fig:nf_allowed_complex_c1_2}) turns out to be very interesting. Firstly, the combination
of inclusive and $B_d \to J/\psi K_S$ constraints only leaves a few small regions of the complex $C_1$ plane where agreement is obtained
between the full compliment of observables and their respective experimental averages, including some where
$C_1$ has an imaginary part of close to \(\pm 0.2\).  Secondly, whereas naive factorisation is not expected to well describe class II colour
suppressed decays, we find that in one of the allowed regions with complex $C_1$, \(r_{21}\) happens to be close to its NF prediction.
This requires a small negative imaginary BSM shift $ \sim -0.2 i$ to \(C_1\).
When considering \(\Delta C_2\), we again observe a strong complementarity of the constraints.
A broad band centred on real shifts is compatible with $B_d \to J/\psi K_S$ data, as well as a diagonal region with negative real and imaginary parts.
Unfortunately the other constraints we consider have no clear region of overlap where all the predictions can be brought into agreement with data,
due to a mild tension between the lifetime ratio on the one hand and radiative decay and the $B_s$ width difference on the other hand.
When considering  \(\Delta C_3\) and \(\Delta C_4\), there are too many non-perturbative parameters in play to obtain constraints from
$B_d \to J/\psi K_S$. 

Turning now to our results for \(\Delta C^{(\prime)}_{5-10}\) and \(\Delta C^{ \prime}_{1-4}\), we group them into three categories exhibiting similar behaviour.
Consider \(\Delta C_{1-4}^{\prime}\), it was found that the strongest constraint upon these coefficients comes indirectly from angular observables through our displayed contours of constant $C^{\prime\mathrm{eff,BSM}}_{9}$ at the fitted values of $C^{\prime\mathrm{eff,BSM,exp}}_{9}\pm1\sigma$, and to a lesser degree, contours of constant $\bar{C}^{\prime\mathrm{eff}}_{7\gamma}$  at the fitted values of $\bar C^{\prime\mathrm{eff,exp}}_{7\gamma}\pm1\sigma$. In all panels we would expect viable values of pairs of BSM coefficients to lie within the region bounded by these contours. We obtain no strong constraint from radiative decay for these coefficients as the small 2 loop mixing of  $C^{\prime\mathrm{eff}}_{7\gamma}$ with this set of coefficients in addition to the purely quadratic dependence of the branching fraction upon $\bar{C}^{\prime\mathrm{eff}}_{7\gamma}$, leads to relaxed bounds. Of all the scenarios considered, we find that pairs involving \(\Delta C^{\prime}_{1} - \Delta C^{\prime}_{3}\) and \(\Delta C^{\prime}_{1} - \Delta C^{\prime}_{4}\) stand out as scenarios where agreement with all data can be found.
Considering now \(\Delta C_{5-10}\), in contrast to the above case, the mixing of $C^{\prime\mathrm{eff}}_{7\gamma}$  with  \(\Delta C_{5-10}\) occurs at 1-loop and results in these coefficients being very highly constrained by radiative decay, and this indicates that models involving combinations of these coefficients are disfavoured by our study. Finally, we consider the coefficients \(\Delta C^{\prime}_{5-10}\). These are constrained in a similar fashion to their unprimed counterparts by each of our observables except the radiative decay branching ratio, which we replace by contours of $\bar{C}^{\prime\mathrm{eff}}_{7\gamma}=\bar C^{\prime\mathrm{eff, exp}}_{7\gamma}\pm1\sigma$. 
Graphical representations of the allowed parameter space are shown in Figures~\ref{fig:reunprimed}--\ref{fig:reprimed}.

As a step towards converting our constraints into statements on more definite NP models, we considered what the equivalent NP scale $\Lambda_\text{NP}$ we are probing when we place limits on our Wilson coefficients, and our results were shown in Table~\ref{tab:lambdas}.
The tensor operators \(Q^{(\prime)c}_9\) are sensitive to the highest scales, with the best fit to those coefficients corresponding to scales in excess of \SI{10}{\TeV}.
All our operators probe scales above \SI{2}{\TeV}, showing how our choice of observables can complement direct LHC searches for NP effects.

\acknowledgments
We thank A.\ Rosov for providing numerical values for the $B \to K$ form factor. The authors are grateful to the Mainz Institute for Theoretical Physics (MITP) for its hospitality and its partial support during the completion of this work.
M.K.\ was supported by MIUR (Italy) under a contract PRIN 2015P5SBHT and by INFN Sezione di Roma La Sapienza and partially supported by the ERC-2010 DaMESyFla Grant Agreement Number: 267985 and an IPPP STFC studentship. A.L.\ is supported by the STFC grant of the IPPP. S.J.\ is supported in past by UK STFC Consolidated Grant ST/P000819/1. K.L.\ acknowledges
support from a PhD studentship jointly funded by STFC and the School of Mathematical and Physical Sciences of the University of Sussex.
The whole project was supported by an IPPP Associateship.
We thank A.\ Rusov and M.L.\ Piscopo for bringing the missing contributions in \eqref{eq:gamma12bsm} (those proportional to $B'_{4,5}$) in a previous version of this paper to our attention during their work on \cite{Lenz:2022pgw}, and for their thorough checking of our other results.

\appendix

\section{Explicit expressions for anomalous dimensions matrices}
\label{app:adm}

Note that as mentioned in Section~\ref{sub:remarks_adm}, the two-loop mixing of \(Q_{3,4}^c\) into the gluon penguin \(Q_{8g}\) has not been calculated -- this corresponds to the zeros in the third and fourth elements of \(\vec{\gamma}_{A8}^{(0)}\).

\allowdisplaybreaks
\begin{align}
\hat{\gamma}_{AA} &= \frac{\alpha_s}{4\pi}
\begin{pmatrix}
  -2&6&0&0&0&0&\\
  6&-2&0&0&0&0\\
  0&0&2&-6&0&0\\
  0&0&0&-16&0&0\\
  0&0&0&0&-16&0\\
  0&0&0&0&-6&2\\
\end{pmatrix}
\\
\hat{\gamma}_{BB} &= \frac{\alpha_s}{4\pi}
\begin{pmatrix}
  2&-6&-\frac{7}{6}&-\frac{1}{2}\\
  0&-16&-1&\frac{1}{3}\\
  -56&-24&-\frac{38}{3}&6\\
  -48&16&0&\frac{16}{3}\\
\end{pmatrix}
\\
\hat{\gamma}_{Ap} &= \frac{\alpha_s}{4\pi}
\begin{pmatrix}
  0&0&0&0\\
  0&\frac{4}{3}&0&0\\
  0&0&0&0\\
  0&-\frac{2}{3}&0&0\\
  0&0&0&0\\
  0&0&0&0\\
\end{pmatrix}
\\
\hat{\gamma}_{pp} &= \frac{\alpha_s}{4\pi}
\begin{pmatrix}
  0&-\frac{52}{3}&0&2\\ 
  -\frac{40}{9}&-\frac{100}{9}&\frac{4}{9}&\frac{5}{6}\\ 
  0&-\frac{256}{3}&0&20\\ 
  -\frac{256}{9}&\frac{56}{9}&\frac{40}{9}&-\frac{2}{3}\\ 
\end{pmatrix}
\\
\gamma_{77} &= \frac{\alpha_s}{4\pi} \left( \frac{32}{3} \right)
\\
\gamma_{87} &= \frac{\alpha_s}{4\pi} \left( -\frac{32}{9} \right)
\\
\gamma_{88} &= \frac{\alpha_s}{4\pi} \left( \frac{28}{3} \right)
\\
\vec{\gamma}_{A7} &= \frac{\alpha_s}{4\pi} 
\begin{pmatrix}
  0 & \frac{464}{81} & 0 & \frac{200}{81} & 0 & 0
\end{pmatrix}^T
\\
\vec{\gamma}_{A8} &= \frac{\alpha_s}{4\pi}
\begin{pmatrix}
  3 & \frac{76}{27} & 0 & 0 & 0 & 0
\end{pmatrix}^T
\\
\vec{\gamma}_{p7} &= \frac{\alpha_s}{4\pi}
\begin{pmatrix}
  \frac{64}{81} & -\frac{200}{243} & -\frac{6464}{81} & -\frac{11408}{243}
\end{pmatrix}^T
\\
\vec{\gamma}_{p8} &= \frac{\alpha_s}{4\pi}
\begin{pmatrix}
  \frac{368}{27} & -\frac{1409}{162} & \frac{13052}{27} & -\frac{2740}{81}
\end{pmatrix}^T
\\
\vec{\gamma}_{p9} &= \frac{\alpha_s}{4\pi}
\begin{pmatrix}
  -\frac{16}{9} &\frac{32}{27} & -\frac{112}{9} & \frac{512}{27}
\end{pmatrix}^T
\\ 
\vec{\gamma}_{B7} &= 
\begin{pmatrix}
  \frac{2m_c}{m_b} &\frac{2m_c}{3m_b} &-\frac{8m_c}{m_b} &-\frac{8m_c}{3m_b}
\end{pmatrix}^T
\\
\vec{\gamma}_{B8} &= 
\begin{pmatrix}
  0 & \frac{m_c}{m_b} & 0 & -\frac{4 m_c}{m_b}
\end{pmatrix}^T
\\
\vec{\gamma}_{A9} &=
\begin{pmatrix}
  -\frac{8}{3} &-\frac{8}{9} &\frac{4}{3} &\frac{4}{9} 
\end{pmatrix}^T
\end{align}

\bibliographystyle{JHEP}
\bibliography{refs}

\end{document}